\documentclass[a4paper,11pt]{article}

\usepackage{amsmath,epsfig}
\usepackage{subfigure}
\usepackage{amssymb,amsfonts}
\usepackage{latexsym}
\usepackage{epsfig}

\usepackage{braket}
\usepackage{amsmath}
\usepackage{amssymb}
\usepackage{color}
\usepackage{cite}
\usepackage{pdfsync}

\makeatletter
\@addtoreset{equation}{section}
\renewcommand{\theequation}{\thesection.\@arabic\c@equation}
\makeatother
\makeatletter
\renewcommand\appendix{\par
  \setcounter{section}{0}%
  \setcounter{subsection}{0}%
  \gdef\thesection{Appendix \@Alph\c@section }
  \renewcommand{\theequation}
  {\Alph{section}.\arabic{equation}}
}
\makeatother

\def \be {\begin{equation}}
\def \ee {\end{equation}}
\def \ba {\begin{array}}
\def \ea {\end{array}}
\def \bea{\begin{eqnarray}}
\def \eea{\end{eqnarray}}

\title{\textbf{{Note on stability of new hyperbolic AdS black holes and phase transitions in R\'{e}nyi entropies}}}
\author{
{Zhen Fang,}$^{1}$\footnote{fangzhen@itp.ac.cn}\,
{Song He,}$^{2,3,1}$\footnote{hesong17@gmail.com}\,
{Danning Li}$^{4,1}$\footnote{lidn@itp.ac.cn}}
\date{}

\begin{document}

\maketitle

\begin{center}
{\it
$^{1}${CAS Key Laboratory of Theoretical Physics, Institute of Theoretical Physics, Chinese Academy of Sciences,
Beijing 100190, P. R. China }\\
\vspace{2mm}
$^{2}${Max Planck Institute for Gravitational Physics (Albert Einstein Institute)
Am M\"{u}hlenberg 1, 14476 Golm, Germany}\\
$^{3}${Yukawa Institute for Theoretical Physics, Kyoto University, Kitashirakawa Oiwakecho,\\
Sakyo-ku, Kyoto 606-8502, Japan}\\
$^{4}${Department of Physics and Siyuan Laboratory, Jinan University, Guangzhou 510632, P.R. China}\\
}
\vspace{10mm}
\end{center}

\begin{abstract}
We construct a series of new hyperbolic black hole
solutions in Einstein-Scalar system  and we apply holographic approach to  investigate the spherical R\'{e}nyi entropy in various deformations {of dual conformal field theories (CFTs)}. Especially, we introduce various
powers {of scalars} in the scalar potentials {for massive and massless scalar}. These scalar potentials correspond to deformation of dual CFTs.
Then we solve {asymptotically} hyperbolic AdS black hole solutions numerically. {We map the instabilities of these black hole solutions to phase transitions of field theory in terms of CHM mapping between hyperbolic hairy AdS black hole and spherical R\'{e}nyi entropy in dual field theories.} Based on these solutions, we study the temperature dependent condensation of dual operator of massive and massless scalar respectively. These condensations show that there might {exist} phase transitions in {dual} deformed CFTs. We also compare free energy between {asymptotically hyperbolic AdS black hole solutions} and hyperbolic {AdS Schwarz (AdS-SW) black hole} to {test} phase transitions. In order to confirm the existence of phase transitions, we turn on linear in-homogenous perturbation to test stability of these hyperbolic {hairy} AdS black holes. In this paper, we show how potential parameters affect the stability of hyperbolic black holes in  several specific examples. For {general values of} potential parameters, it needs further studies to see how the transition happens. Finally, we comment on these instabilities associated with spherical R\'{e}nyi entropy in dual deformed CFTs.
\end{abstract}

\baselineskip 18pt
\thispagestyle{empty}

\newpage

\section{Introduction}
The stability of black holes in anti-de Sitter space has been widely studied in the context of the AdS/CFT
correspondence \cite{Maldacena:1997re}\cite{Gubser:1998bc}\cite{Witten:1998qj}\cite{Aharony:1999ti}. {The investigation of thermodynamical stability of black hole provides a novel window on the phase structure of the dual CFTs.} In holographic approaches to condensed matter physics the
instability of a black hole {due} to the condensation of scalar hair is dual to a superconducting
phase transition \cite{Hartnoll:2008vx}\cite{Hartnoll:2008kx}. The physical relevance is also related to {studying} phase transitions in AdS/QCD literature \cite{Erlich:2005qh}\cite{Gursoy:2007cb}\cite{Gursoy:2007er}\cite{Gubser:2008ny}\cite{Li:2011hp}. {These phase transitions correspond to freezing or releasing the degree of freedom in such systems. Entanglement entropy (EE) can measure the effective degree of freedom in quantum system.} In gravity side, entanglement entropy can be calculated by RT formula \cite{Ryu:2006bv}\cite{Ryu:2006ef}. In holographic condensed matter literature, those phase transitions can {also be} confirmed by entanglement entropy or R\'{e}nyi entropy by AdS/CFT, for various examples, \cite{Klebanov:2007ws}\cite{Albash:2012pd} \cite{Cai:2012sk}\cite{Cai:2012nm}\cite{Belin:2013dva}\cite{Belin:2014mva}. Therefore, one can calculate entanglement entropy or {entanglement R\'{e}nyi entropy (ERE)} in gravity side to test phase transition in field theory.

{The standard approach to calculate entanglement entropy in field theory is called replica trick \cite{Callan:1994py}\cite{Calabrese:2004eu}\cite{Calabrese:2005zw}. Recently, the replica trick has been also applied in gravity side \cite{Lewkowycz:2013nqa}\cite{Song:2016pwx} to confirm holographic dictionary of EE \cite{Ryu:2006bv}\cite{Ryu:2006ef}. The ERE for {vacuum states} in various situations \cite{Klebanov:2011uf}\cite{Nishioka:2013haa}\cite{Agon:2013iva}\cite{Huang:2014gca}\cite{Hama:2014iea} has been studied. More recently, the ERE for local excited states in CFTs have been extensively investigated in \cite{Nozaki:2014hna}\cite{Nozaki:2014uaa} \cite{He:2014mwa}\cite{Caputa:2014vaa}\cite{Guo:2015uwa}\cite{Chen:2015usa}\cite{Nozaki:2015mca}\cite{Sarosi:2016oks}\cite{Sarosi:2016atx}. In string theory, \cite{He:2014gva} tried to use replica trick to calculate entanglement entropy associated {with} black hole entropy. From holography, \cite{Dong:2016fnf} proposed a general framework to study {ERE. The related} shape deformations of ERE have {also been} intensively studied in \cite{Balakrishnan:2016ttg}\cite{Bianchi:2016xvf}\cite{Chu:2016tps} recently.}

We will {firstly} focus on the instability of hyperbolic AdS black holes and finally comment on holographic spherical R\'{e}nyi entropy. For spherical R\'{e}nyi entropy of the ground state in CFTs, \cite{Casini:2011kv}\cite{Hung:2011nu} proposed that it is equal to thermal entropy of higher dimensional hyperbolic AdS black hole by so called CHM mapping {method} \cite{Casini:2011kv}\cite{Hung:2011nu}\cite{Myers:2010xs}\cite{Myers:2010tj}. {The main goal of the present paper is to study how to make use of this dictionary to test the phase structures in deformed field theory side. The relevant deformations in field theory correspond to adding massive scalar or massless scalar in the dual hyperbolic black hole.}
{We review how hyperbolic AdS black holes relate to holographic R\'{e}nyi entropy. A holographic calculation of R\'{e}nyi entropy for a
spherical entangling surface is derived in \cite{Casini:2011kv}\cite{Hung:2011nu}\cite{Myers:2010xs}\cite{Myers:2010tj}. Applying this approach, there are many {extended} studies \cite{Belin:2014mva}\cite{Belin:2013uta}\cite{Brown:2015bva}. Following CHM mapping, the
density matrix is thermal and we can write the $n$'th
power of $\rho$ as following}
 \bea\label{8.1}
 \rho^{n}=U^{-1}\,{\exp\left[-n H/T_0\right]\over Z(T_0)^n}\,U \quad{\rm
 with}\ \ Z(T_0)\equiv \textrm{tr}\left[e^{-H/T_0}\right]\,,
 \eea where $n$ is {an} integer number.
The unitary transformation $U$ and its inverse will be
canceled {by} taking the trace of this expression. Hence the trace of the $n$'th power of density matrix is
 \bea\label{8.2}
\textrm{tr}\left[\,\rho^{\,n}\right] = {Z(T_0/n)\over Z(T_0)^n}.
 \eea
{Using} the definition of the free energy of dual black hole, i.e. $F(T)=-T\log
Z(T)$, the corresponding R\'{e}nyi entropy becomes
 \bea
 S_n ={n\over(1-n)T_0}\left( F(T_0)-F(T_0/n)\right).
 \eea
in terms of {the derivation in} \cite{Casini:2011kv}\cite{Myers:2010xs}\cite{Myers:2010tj}.
Further using $S=-\partial F/\partial T$, this
expression can be rewritten as
 \bea
S_n = {n\over n-1} {1\over T_0}\, \int_{T_0/n}^{T_0}
S_\text{therm}(T)\, dT\,,
 \label{finalfor111}
 \eea
where $S_n$ is the R\'{e}nyi entropy while $S_{\text{thermal}}(T)$ denotes
the thermal entropy of the CFT on $R\times H^{d-1}$. The entanglement entropy can be
 \bea
 S_{\text{EE}}=\lim_{n\to 1}S_n = S_\text{thermal}(T_0)\,.
 \label{lim1}
 \eea
with $T_0$ given by $ {1\over 2\pi R}$. Here, $R$ is the curvature scale on the hyperbolic spatial slices $H^{d-1}$ matching the radius of the original spherical entangle surface, $R$. {This proposal gives us a way to connect the hyperbolic AdS black holes with ERE. }

{There are several hyperbolic AdS black holes, which have been
constructed in} \cite{Aminneborg:1996iz}\cite{Mann:1996gj}\cite{Mann:1997zn}\cite{Emparan:1999gf}\cite{Emparan:1998he}\cite{Birmingham:1998nr}. In \cite{Dias:2010ma},
it was shown that static black holes with hyperbolic horizons can become unstable to the formation of uncharged scalar hair on the horizon of the black hole due to the presence of an extremal limit with near-horizon geometry $AdS_2 \times H^3$  \cite{Robinson59}\cite{Bertotti59}\cite{Bardeen:1999px}\cite{Brihaye:2011hm}. Furthermore, authors of \cite{Belin:2014lea} introduced a topological black hole with a minimal coupled scalar field {with} negative mass-square and showed this new {instability} appeared. In \cite{Belin:2013dva}, {the authors} mapped the instability of this gravity solution to the phase transition happened in dual CFTs by {CHM mapping.} In \cite{Belin:2014mva}, they investigated charged hyperbolic black holes, which became unstable to presence of scalar hair at sufficiently low temperature. Such kind of instability is the same as the holographic superconducting instability in boundary hyperbolic space. {In summary, scalar fields with masses below the effective Breitenlohner-Freedman bound for the near-horizon $AdS_2$ will induce instability at sufficient low temperatures. This happens for both charged and uncharged black holes.}

{In this paper, we will construct a series of general hyperbolic AdS black holes with neutra scalar.} More precisely, in this system, we introduce series of specific powers of scalar in scalar potential. In \cite{Belin:2013dva}, the authors showed that there was an instability in massive scalar hariy hyperbolic AdS black hole. The instability {might} induce a phase transition and {study on} entanglement R\'{e}nyi entropy also confirmed the phase transition. {In our setup, we extend the studies in \cite{Belin:2013dva} by introducing higher powers of scalar self-interactions to deform dual CFTs. We start with a general gravity setup and see what will happen. Firstly, we work out these gravity {solutions} in UV region. Secondly, we can find hyperbolic AdS black hole solution numerically in various scalar potentials. In terms of CHM mapping, basing on these hyperbolic AdS hairy black holes, we obtain the spherical ERE in dual deformed CFTs. ERE obtained in our setup shows that there exist phase transitions in dual CFTs.} We can extract the condensation of dual operator with respect to temperature in each solution. The condensation of dual operator {indicates} that the phase transition might happen. {To confirm} the phase transitions, we compare the free energy between the hyperbolic scalar hairy AdS black hole solutions (HSHAdS) and hyperbolic AdS-SW black hole to reveal the transition. {Furthermore}, we turn on the in-homogenous linear perturbation to test the stability of HSHAdS and the stability condition highly constrains the potential parameters presented in the massive and massless scalar potential. We will give some explicit examples to show what kinds of scalar potential will give stable HSHAdS. {blue}{Finally, one can make use of the stability to obtain the phase structure of the dual theories.}

An overview of the remainder of the paper is as follows: in section 2, we firstly set up the gravity which is our starting point. In section 3, we study the boundary energy momentum tensor of hyperbolic hairy black hole solutions with introducing various of boundary counter terms in massless and massive scalars respectively. Furthermore, we evaluate the free energy of these solutions. In section 4, through the above numerical analysis, we found that there are interesting phase transitions in deformed CFTs. We make use of condensation of dual operators and free energy of each solution to {check whether} phase transition will really happen in deformed CFTs {or not}. In section 5, we will demonstrate that the hyperbolic black holes are unstable and entanglement R\'{e}nyi entropies show a phase transition. Therefore, in section 6, we turn to the physical case of these models which are normalizable on hyperboloid. We will devote section 7 to conclusions and discussions.  In \ref{Appendix8}, we will list the asymptotic AdS boundary behavior which is controlled by Einstein equations for massless and massive scalar respectively. These UV behaviors are useful to obtain the numerical solutions {and we also list how to obtain the vacuum expectation value of dual operators}. In \ref{Appendix9}, we show various new hyperbolic scalar hairy AdS black hole solutions numerically as examples to {check the validity of our numerical procedure}.

\section{Gravity Setup}\label{gravity-setup}
{The action of Einstein-Dilaton system} in 5D spacetime in Einstein frame is
\begin{equation} \label{minimal-Einstein-action}
S_{5D}=\frac{1}{16 \pi G_5} \int d^5 x
\sqrt{-g}\left(R-\frac{4}{3}\partial_{\mu}\phi\partial^{\mu}\phi-V(\phi)\right).
\end{equation}
Here $G_5$ is the 5D Newton constant, $g$ is the 5D metric determinant and $\phi,V$ are the scalar field and the corresponding potential. In this paper, we study the potential {of the form} $V=\frac{1}{L^2}\Big(-12+v_2\phi^2+v_3 \phi ^3+v_4 \phi ^4+v_6 \phi ^6\Big)$ for simplification. From these cases, one can learn how the self interaction in the bulk involve in boundary phase structure. In {general} scalar potential, our calculations involve in examining
the Einstein and scalar field equations together and finding solutions where the
scalar has a nontrivial profile reflecting the presence of the relevant deformations
in the boundary theory.

The equations of motion are
\begin{eqnarray} \label{EOM}
E_{\mu\nu}+\frac{1}{2}g_{\mu\nu}\left(\frac{4}{3}
\partial_{\mu}\phi\partial^{\mu}\phi+V(\phi)\right)
-\frac{4}{3}\partial_{\mu}\phi\partial_{\nu}\phi =0,
\end{eqnarray}
where $E_{\mu\nu}=R_{\mu\nu}-\frac{1}{2}Rg_{\mu\nu}$ is Einstein
tensor.

We would like to choose the following ansatz to solve the Einstein equations of motion,
\bea\label{ansatz}
ds^2&=&-\frac{L^2 e^{2 A_e(z)}}{z^2}\left(-f(z) dt^2+\frac{1}{f(z)}dz^2+\left(d\psi^2+\sinh ^2(\psi )d\theta^2+\sin ^2(\theta ) \sinh ^2(\psi )d\varphi^2\right)\right)\nonumber\\
&=&-\frac{L^2 e^{2 A_e(z)}}{z^2}\left(-f(z) dt^2+\frac{1}{f(z)}dz^2+d H^3)\right),
\eea where $H^3$ is 3 dimensional hyperbolic space and $L$ is AdS radius.
In terms of the above ansatz, one can obtain equations,
\bea \label{equations}
A_e''(z)-A_e'(z){}^2+\frac{2 A_e'(z)}{z}+\frac{4}{9} \phi '(z)^2&=&0,\nonumber\\
f''(z)+f'(z) \left(3 A_e'(z)-\frac{3}{z}\right)-\frac{4}{L^2}&=&0,\nonumber\\
\phi ''(z)+ \left(3 A_e'(z)+\frac{f'(z)}{f(z)}-\frac{3}{z}\right)\phi '(z)-\frac{3 L^2 e^{2 A_e(z)} V'(\phi (z))}{8 z^2 f(z)}&=&0.\label{eq-phi}
\eea
One more constrain equation is
\bea\label{constrain}
6 A_e'(z){}^2+\left(\frac{3 f'(z)}{2 f(z)}-\frac{12}{z}\right) A_e'(z)+\frac{L^2 e^{2 A_e(z)} V(\phi (z))}{2 z^2 f(z)}-\frac{3 f'(z)}{2 z f(z)}+\frac{3}{L^2f(z)}+\frac{6}{z^2}-\frac{2}{3} \phi '(z)^2=0\nonumber\\
\eea
(\ref{constrain}) is not independent on the other three equations in (\ref{equations}). Once the gravity solution is obtained from (\ref{equations}), one could use (\ref{constrain}) to check the solution.

Here, we note that (\ref{eq-phi}) would impose a natural boundary condition near horizon. If one collects all the terms with a denominator $f(z)$, the results are as following
\begin{eqnarray}
\frac{Q(z)}{8z^2f(z)}
\end{eqnarray}
with $Q(z)\equiv 8z^2f^{'}\phi^{'}-3L^2e^{2A_e}V^{'}(\phi)$. Since the horizon is not a real singularity, the apparent singularity $f(z_h)=0$ in Eq.(\ref{eq-phi}) should be canceled by requiring $Q(z_h)=0$. Later, we will try to solve this boundary value problem using numerical method {described in appendix A and developed in Ref.\cite{He:2013rsa}}. In this paper, we show some details in \ref{Appendix8} and \ref{Appendix9}. In numerical procedure, we set $G=L=1$ to simplify our numerical calculation.

\section{Energy Momentum Tensor and Free energy}
In this section, we turn to study the stability of hyperbolic AdS black hole solutions.
We only focus on two cases. {The one} is massless scalar case and the other is massive case. The massive and massless neutral scalar correspond to specific QCD operator, e.g. dimension 4 glueball operator $\langle O_1 \rangle$ and dimension 2 glueball operator $\langle O_2\rangle$ \footnote{{We have shown how to read out the $\langle O_1 \rangle$ and $\langle O_2 \rangle$ in series expansion of scalar near the UV region in appendix \ref{app-a2} and \ref{app-a1} respectively.}} respectively in this paper. {Our studies will be helpful to understand how deconfinement transition from holographic point of view.} The {R\'{e}nyi} entropy is very good quantity to mimic these phase transitions in QCD literature. Furthermore, we extend CHM transformation to calculate R\'{e}nyi entropy in sub classes of non-conformal theories. {These non-conformal theories are obtained by adding simple deformations operators\footnote{These deformations only depend on holographic direction, namely $z$ in our paper.}} which correspond to neutral scalar with self interaction potential.

To obtain well defined energy
momentum tensor on the boundary, one should introduce the suitable
counter terms. For later use, we will work out a well defined counter term for these gravity solutions and these terms will
be also used in studying free energy and spherical R\'{e}nyi entropy of dual CFTs.

\subsection{Energy Momentum Tensor}
In this subsection, we would like to introduce the counter terms to
cancel the UV divergences of the {on-shell action and make the energy momentum
tensor of dual field theory well defined. Firstly, we introduce general gauge invariant counter terms with undetermined coefficients in our system. Finally, we can solve these coefficients to cancel the divergences in massless and massive cases respectively in this paper.}
\subsubsection{Massless Scalar Cases}
For massless scalar case, the total action now becomes
\begin{eqnarray}\label{C2}
I_{\text{ren}}&=&S_{\text{5D}}+
S_{\text{GH}}+S_{\text{count}}\nonumber\\&=&\frac{1}{16 \pi G_5}
\int_M d^5
x\sqrt{-g}\left(R-\frac{4}{3}\partial_{\mu}\phi\partial^{\mu}\phi-V(\phi)\right)\nonumber \\&-&
\frac{1}{16\pi G_5} \int_{\partial
M}d^4x\sqrt{-\gamma}\Big[2K-{6\over L}+ \lambda_1 \mathcal{R}+ \lambda_2 \mathcal{R}_{ab}\mathcal{R}^{ab}+\lambda_3 \mathcal{R}^2+... \Big],\nonumber \\
\end{eqnarray}
with $\lambda_1, \lambda_2, \lambda_3$ undermined coefficients of counter
terms \cite{Henningson:1998gx} \cite{Henningson:1998ey}\cite{Balasubramanian:1999re}\cite{Hyun:1998vg}\cite{Emparan:1999pm}\cite{Mann:1999pc}  $\mathcal{R}$, $\mathcal{R}_{ab}\mathcal{R}^{ab}, \mathcal{R}^2 $   to be worked out later. Here we choose massless scalar potential as
$V=\frac{1}{L^2}\Big(-12+v_3 \phi ^3+v_4 \phi ^4+v_6 \phi ^6\Big)$ and $v_3,v_4,v_6$ are free parameters.
The first term of the last line in (\ref{C2}) is Gibbons-Hawking
term $S_{\text{GH}}$ and the remain terms are $S_{\text{count}}$ related to cosmological constant and scalar
field. These coefficients
can be fixed by canceling the divergences of boundary momentum
tensor. Here $K_{ij}$\footnote{We use index $\mu,\nu$ and $i,j$ to denote bulk coordinates and the boundary coordinates respectively.} and $K$ are respectively the extrinsic
curvature and its trace of the boundary $\partial M$, $\gamma_{ij}$
is the induced metric on the boundary $\partial M$. These quantities
are defined as follows
\begin{eqnarray}
\gamma_{\mu \nu}&=&g_{\mu \nu} +n_{\mu} n_{\nu},\\
K_{\mu\nu}&=&h^\lambda_\nu D_{\lambda} n_{\mu},\\
\gamma&=&\det(\gamma_{\mu\nu}),\\
K&=&g^{\mu\nu}K_{\mu\nu},
\end{eqnarray}
where $\gamma_{\mu \nu}$ denotes the induced metric, $n_{\mu}$
stands for the normal direction to the boundary surface $\partial M$
as well as $D_{\lambda}$ stands for covariant derivative. Finally, $\mathcal{R}$ and $\mathcal{R}_{ab}$ are the Ricci scalar and Ricci tensor for the
boundary metric respectively. In general cases, one should introduce higher powers of $\mathcal{R}$ and  various combination of $\mathcal{R}_{ab}$ to cancel
the total UV divergence. For massive and massless cases in this paper, we just introduce $\mathcal{R}$ to cancel all the UV divergence. That means we can
set $\lambda_2, \lambda_3$ to be vanishing.

In the asymptotical AdS hyperbolic black hole, the boundary surface locates at $z=0$
surface, and usually one has to regularized it to a finite
$z=\epsilon$ surface. So we have the normalized normal vector
$n_\mu=\frac{\delta^\mu_z}{\sqrt{g_{zz}}}$.

To regulate the theory, we restrict to the region $z\ge \epsilon$
and the surface term is evaluated at $z=\epsilon$. The induced
metric is $\gamma_{ij}=\frac{\tilde{L}^2}{\epsilon^2}
g_{ij}(x,\epsilon)$, where the leading term of expansion of
$g_{ij}(x,\epsilon)$ with respect to $\epsilon$ is the flat metric
$g_{(0)}^{ij}$. Then the one point function of stress-energy tensor
of the dual CFT is given by \cite{KS}\cite{SKS}\cite{Nojiri:1998dh}\cite{Nojiri:2000kh}
\begin{eqnarray}
T_{ij}&=&\frac{2}{\sqrt{-\det g_{(0)}}}\frac{\delta I_{ren}}{\delta
g_{(0)}^{ij}}=\lim_{\epsilon \to
0}\Big(\frac{{L}^2}{\epsilon^2}\frac{2}{\sqrt{-\gamma}}\frac{\delta
I_{ren}}{\delta \gamma^{ij}}\Big). \label{CFTET}
\end{eqnarray}
The finite part of boundary
energy-stress tensor is from the
$O(\epsilon^2)$ terms of the Brown-York tensor $T_{ij}$ on the
boundary $z=\epsilon$, with
\begin{eqnarray}\label{BY}
T_{ij}&=&-\frac{1}{16 \pi G_5}\Big[K_{ij}-\Big((K+{d-2\over
L})g_{ij}-\lambda_1 \mathcal{R}_{ij}-2\lambda_2 \mathcal{R}_{ik}\mathcal{R}_{jk}+{\lambda_1\over 2}g_{ij}\mathcal{R}-
2 \lambda_3 \mathcal{R}_{ij}\mathcal{R}\nonumber\\&+&{1\over 2}\lambda_3 g_{ij}\mathcal{R}^2-2 \lambda_2 \mathcal{R}^{kl}\mathcal{R}_{ikjl}+\lambda_2\mathcal{R}_{j}^{k}\mathcal{R}_{il}^{k l}
+\lambda_2\mathcal{R}_{i}^{k}\mathcal{R}_{jl}^{k l}+{\lambda_2\over 2}g_{ij}\mathcal{R}_{kl}^{k m}\mathcal{R}_{ln}^{ mn}\nonumber\\&+& (2 \lambda_3+\lambda_2) \nabla_{j}\nabla_{i}\mathcal{R}-\lambda_2\mathcal{R}_{ij;k}^{;k }
-(2 \lambda_3+{1\over 2}\lambda_2) g_{ij }\mathcal{R}_{kl;m}^{kl; m}\Big)\Big].
\end{eqnarray}

In the massless scalar hair hyperbolic AdS black hole, the coefficients of counter terms can be
following
\begin{eqnarray}
\lambda_1&=&{1\over 2},\nonumber\\
\lambda_2&=&0,\nonumber\\ \lambda_3&=&0,
\end{eqnarray} where we have fixed these coefficients by removing the UV divergence $z\rightarrow 0$ appeared in on-shell action of massless scalar.
Directly evaluate (\ref{BY}) using (\ref{CFTET}), we get $tt$ component of energy momentum tensor \bea
T_{tt}={1\over 16\pi G}\Big(\frac{3L}{8}-\frac{3 f_4 L}{2}\Big). \label{ETGB}\eea

\subsubsection{Massive Scalar Cases}
{For massive scalar, the total action will be different from massless cases. The main reason is that the UV behavior of massive scalar is different from the massless cases. In massive case, we will introduce following counter term to cook up well defined on-shell action.}
\begin{eqnarray}\label{C22}
I_{\text{ren}}&=&S_{\text{5D}}+
S_{\text{GH}}+S_{\text{count}}\nonumber\\&=&\frac{1}{16 \pi G_5}
\int_M d^5
x\sqrt{-g}\left(R-\frac{4}{3}\partial_{\mu}\phi\partial^{\mu}\phi-V(\phi)\right)\nonumber \\&-&
\frac{1}{16\pi G_5} \int_{\partial
M}d^4x\sqrt{-\gamma}\Big[2K-{6\over L}+ \lambda_{m1} \mathcal{R}+ \lambda_{m2}\phi^2 +\lambda_{m3}\phi \mathcal{R}+... \Big],\nonumber \\
\end{eqnarray} Here massive scalar potential are chosen to be $V=\frac{1}{L^2}\Big(-12-\frac{16 \phi ^2}{3}+v_3 \phi ^3+v_4 \phi ^4+v_6 \phi ^6\Big)$ and $v_3,v_4,v_6$ are free parameters.
In terms of (\ref{CFTET}), the boundary energy momentum tensor would be
\bea
T_{ij}&=&{1\over 16\pi G}\Big[ K_{ij}-(K+{d-2\over 2}-\lambda_{m2}\phi^2)g_{ij}+(\lambda_{m3}\phi+\lambda_{m1})(\mathcal{R}_{ij}-{1\over 2}g_{ij}\mathcal{R})\Big]
\eea
In the massive scalar hairy hyperbolic AdS black hole, the coefficients of count terms can be
following\begin{eqnarray}
\lambda_{m1}&=&{1\over 2},\nonumber\\
\lambda_{m2}&=&{8\over 3},\nonumber\\ \lambda_{m3}&=&{2 \langle O_2\rangle\over 9},
\end{eqnarray}
where $\langle O_2\rangle$ corresponds to expectation value of dual operator $O_2$ of massive scalar $\phi$. We have fixed these coefficients by removing the UV divergence $z\rightarrow 0$ appearing in on-shell action of massive scalar.

Directly evaluate (\ref{BY}) using (\ref{CFTET}), we get \bea
T_{tt}={1\over 16\pi G}\Big(-\frac{3 f_4L}{2 L^2}-\frac{\langle O_2\rangle^2 L}{6 }+\frac{3L}{8 }\Big). \label{ETGB}\eea

We have introduced counter terms to make well defined boundary stress tensor. With these counter terms, we can obtain on-shell action which will play an important
role in judging the phases of theories in the coming section.

\subsection{The Difference of Free Energy}
After introducing the counter term to remove the divergence of the action, we can work out
the on-shell action which will be helpful to test the holographic phase structures. Later, we
will also make use of condensation of dual operator to get the flavor of phase transitions.

For massless scalar case, the on-shell action can be
\bea
S_{\text{5D-BH}}={1\over 16\pi G}\Big(\frac{3}{4}-f_4\Big)
\eea

For massive scalar case, the on-shell action can be
\bea
S_{\text{5D-BH}}={1\over 16\pi G}\Big(\frac{3}{4}-f_4-\frac{1}{45} 8 p_2^2-\frac{28 p_{22} p_2}{75}-\frac{416 p_{22}^2}{1125}\Big)\Big|_{p_{22}=0}
\eea
where we have to turn off the source $p_{22}$ to obtain the expectation value of dual operator in vacuum for later use.

To {summarize} this section, we have introduced a self-consistent counter term to obtain the well defined free energy by requiring boundary energy momentum to be finite. Once we obtain the free energy of these black hole solutions, we can study phase structures by comparing free energy in the coming section.

\section{Phase Transitions }
In this section, by calculating condensation of dual operator and free energy, we will study the stability of the hyperbolic AdS black hole solutions obtained from \ref{Appendix8} and \ref{Appendix9}. We will show
temperature dependence behaviors of condensation of operators {$O_1$ and $O_2$, which are dual to massless and massive scalar respectively}. Firstly, we will make use of free energy to study the stability of these new hyperbolic black hole solutions. In this section, we mainly focus on the constant modes in which we do not turn on the in-homogenous perturbation of these solutions. The constant mode means that the field configurations only depend on holographic direction $z$. We would emphasize that this analysis is {preliminary} and later we will turn to go further to check the stability of these solutions {in section \ref{section7}. In section \ref{section7}, we will go back to the phase structures in these theories studied in this section in terms of linear perturbation.}
\subsection{Condensation}
In this subsection, we will figure out all field configurations and extract the condensation of dual operator $O_1, O_2$ of massless or massive scalar field to see what will happen with changing related parameters, for example, temperature and coupling constant of scalar self-interaction. Basically, one can extract the condensation of dual operator by UV expansion of massless and massive scalar shown in Eq.(\ref{sol1phi}) Eq.(\ref{massivescalar}) in terms of AdS/CFT dictionary. The condensation will imply whether there is phase transition or not. Later, we will use free energy to confirm these phase transitions and determine the transition temperature.
\subsubsection{Massless Cases}\label{masslessphasetransition}
We would like to introduce several deformations in massless scalar potential, for example, adding $\phi^3, \phi^4, \phi^6$ terms to the potential. We mainly focus on obtaining condensation of the dual $\Delta=4$ {glueball operator} $O_1$ with respect to temperature.

\begin{figure}[!h]
\begin{center}
\epsfxsize=6.5 cm \epsfysize=5.5 cm \epsfbox{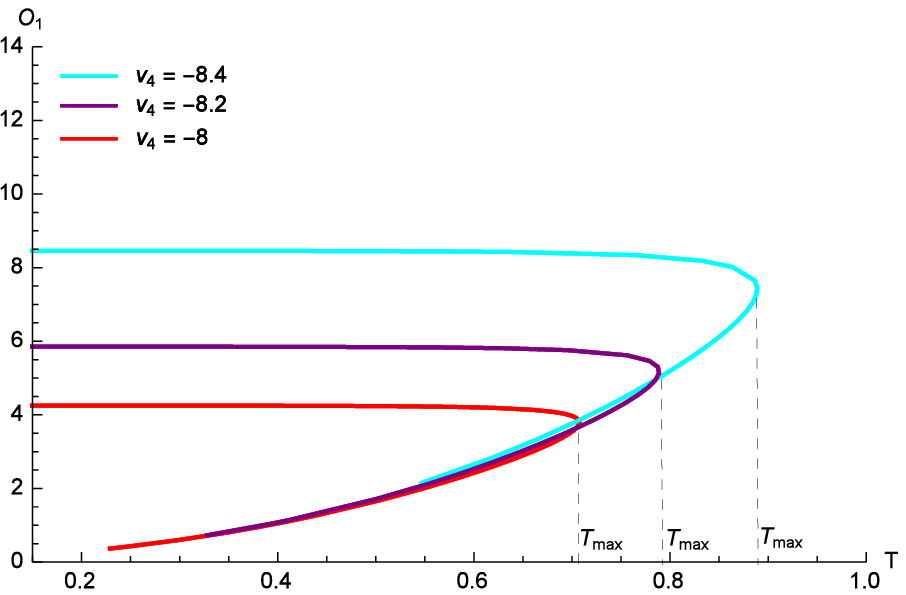}
\hspace*{0.1cm} \epsfxsize=6.5 cm \epsfysize=5.5 cm \epsfbox{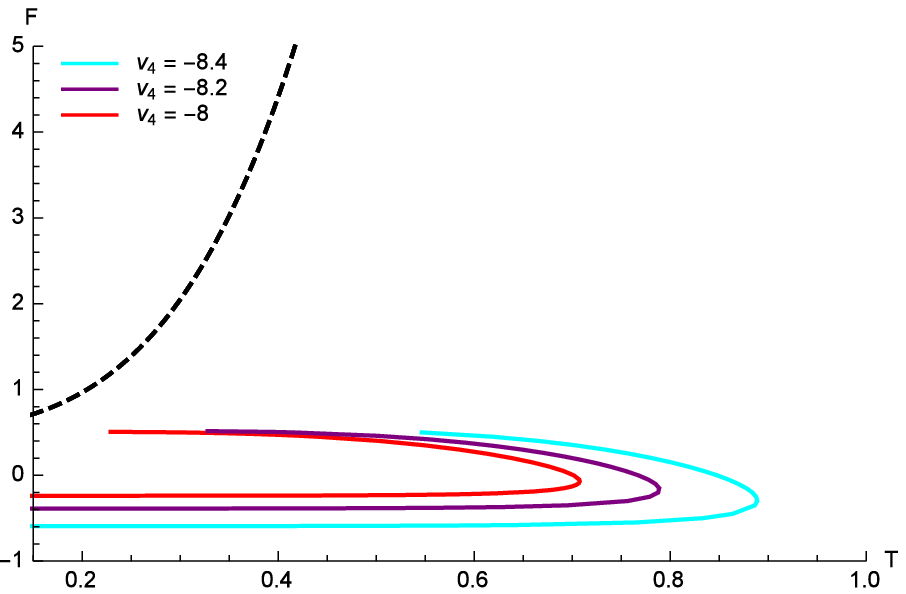}
\hskip 4.5 cm \textbf{( a ) } \hskip 4.5 cm \textbf{( b )}\\
\end{center}
\caption{The condensation is as a function of temperature in massless scalar case with potential $V(\phi )= -{12\over L^2}+{\nu_4\phi ^4\over L^2}$.}\label{zerocondensation-1}
\end{figure}

Firstly, we would like to calculate the condensation in massless scalar with potential like $V(\phi )= -{12\over L^2}+{\nu_4\phi ^4\over L^2}$. In fig.\ref{zerocondensation-1}(a),
we have shown the condensation as a function of temperature. The different colored curves correspond to {choosing different} model parameter $\nu_4$. With increasing $\nu_4$, the condensation at the same temperature will increase gradually. There is a transition temperature {defined where the condensation goes to zero}. For each colored curve, the condensation is
double valued function with respect to temperature from zero temperature to maximal temperature $T_{max}$. In fig.\ref{zerocondensation-1}(b), we calculate free energy with respect to temperature and it shows that the dashed line part is unstable comparing with solid curve. That means the $T_{max}$ is phase transition temperature $T_c$ in terms of free energy. Below the transition temperature $T_c=T_{max}$, the condensation is a monodrome function of temperature. At the transition temperature, the condensation will jump from finite positive value to zero and the massless hairy black hole solution is unstable comparing with hyperbolic AdS-SW black hole. That is to say hyperbolic AdS-SW black hole is favored when $T\geq T_c$. Up to this stage, we find the instability exists in this case.

\begin{figure}[!h]
\begin{center}
\epsfxsize=6.5 cm \epsfysize=5.5 cm \epsfbox{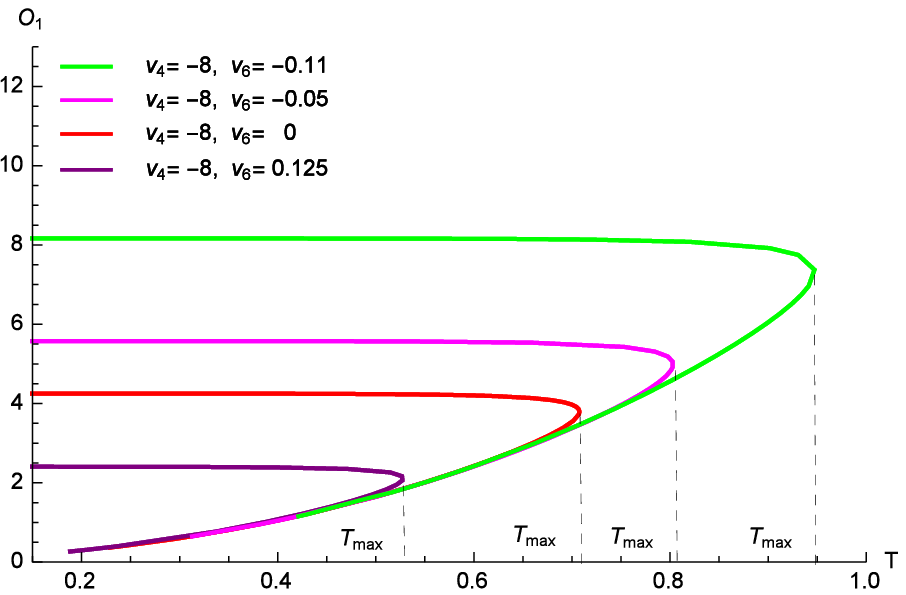}
\hspace*{0.1cm} \epsfxsize=6.5 cm \epsfysize=5.5 cm \epsfbox{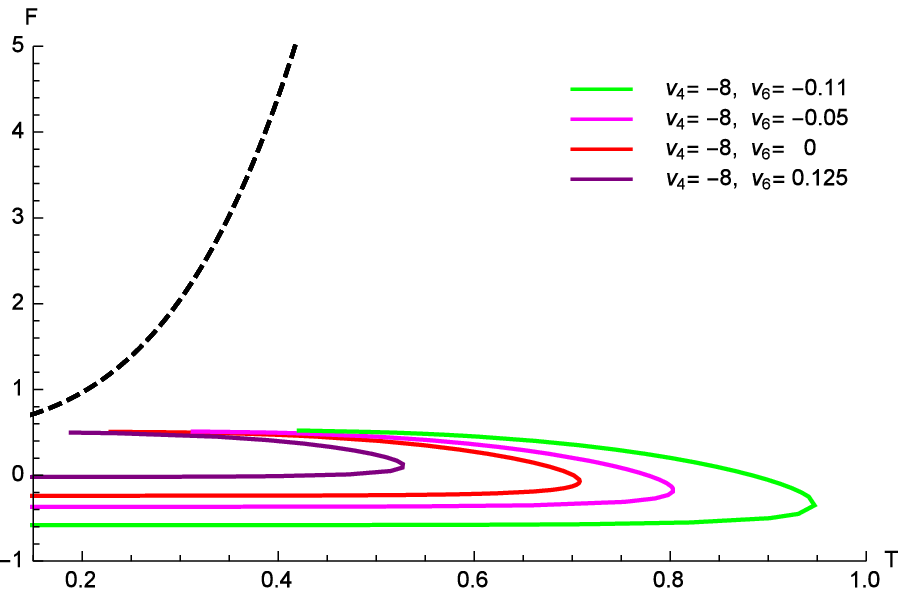}
\hskip 4.5 cm \textbf{( a ) } \hskip 4.5 cm \textbf{( b )}\\
\end{center}
\caption{The condensation is as a function of temperature in massless scalar case with potential $V(\phi )= -{12\over L^2}+{\nu_6\phi ^6\over L^2}$.}\label{zerocondensation-2}
\end{figure}

Secondly, we would like to calculate the condensation $O_1$ in massless scalar with potential like $V(\phi )= -{12\over L^2}+{\nu_4\phi ^4\over L^2}+{\nu_6\phi ^6\over L^2}$.
In fig.\ref{zerocondensation-2}(a),
we have shown the condensation as a function of temperature. The different colored curves correspond to {choosing} different model parameter $\nu_6$ with fixing $\nu_4$. With increasing $\nu_6$, the condensation at same temperature will decrease gradually. There is also a transition temperature {defined where the condensation goes to zero}. For each colored curve, the condensation is double valued function with respect to temperature from zero temperature to maximal temperature $T_{max}$.
 In fig.\ref{zerocondensation-2}(b), we calculate free energy with respect to temperature and it shows that the dashed line part is unstable comparing with solid curve in $T<T_{max}$. That means the $T_{max}$ is phase transition temperature $T_c$. Below the transition temperature $T_c$, the condensation is a monodrome function of temperature. At the transition temperature, the condensation will jump from finite positive value to zero and the massless hairy black hole solution is unstable comparing with hyperbolic AdS-SW black hole in $T>T_{max}$. That is also to say hyperbolic AdS-SW black hole is favored when $T\geq T_c$. Below the transition temperature, the condensation is a monodrome function of temperature. At the transition temperature, the condensation will jump from finite positive value to zero. We can see that ${\nu_6\phi ^6\over L^2}$ does not change the type of phase transition induced by ${\nu_4\phi ^4\over L^2}$.

\begin{figure}[!h]
\begin{center}
\epsfxsize=6.5 cm \epsfysize=5.5 cm \epsfbox{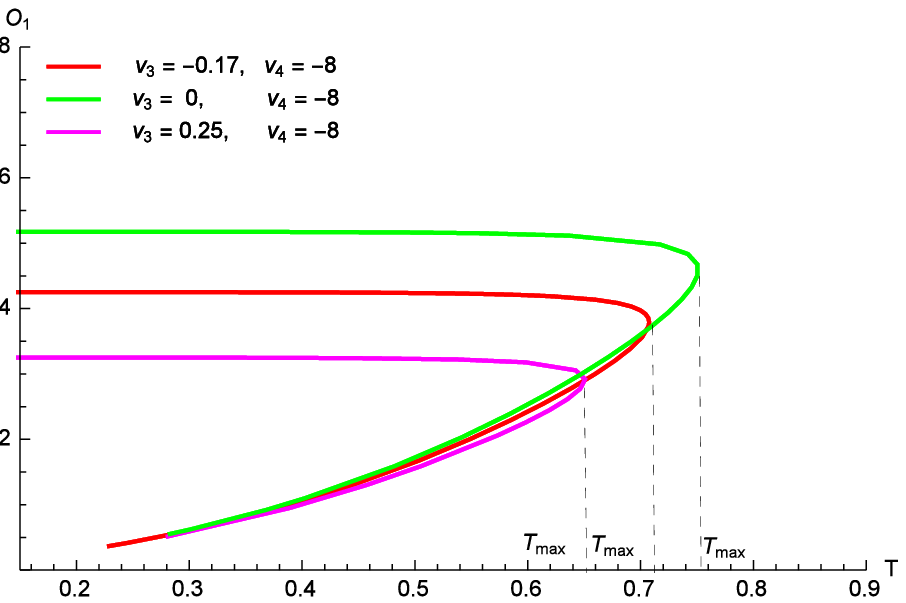}
\hspace*{0.1cm} \epsfxsize=6.5 cm \epsfysize=5.5 cm \epsfbox{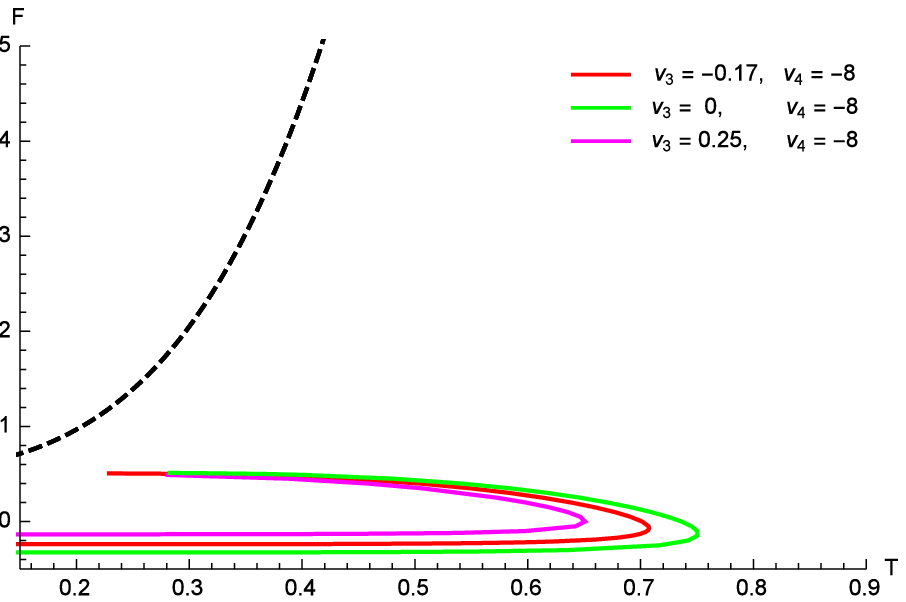}
\hskip 4.5 cm \textbf{( a ) } \hskip 4.5 cm \textbf{( b )}\\
\end{center}
\caption{The condensation is as a function of temperature in massless scalar case with potential $V(\phi )= -{12\over L^2}+{\nu_3\phi ^3\over L^2}+{\nu_4\phi ^4\over L^2}$.}\label{zerocondensation-3}
\end{figure}

Finally, we would like to calculate the condensation in massless scalar with potential like $V(\phi )= -{12\over L^2}+{\nu_3\phi ^3\over L^2}+{\nu_4\phi ^4\over L^2}$.
In fig.\ref{zerocondensation-3}(a), the condensation as a function of temperature has been presented. The different colored curves correspond to {choosing different} model parameter $\nu_3$ with fixing $\nu_4$. With increasing $\nu_3$, the condensation at same temperature will decrease gradually. For each colored curve, the condensation is double valued function with respect to temperature from zero temperature to maximal temperature $T_{max}$.
In fig.\ref{zerocondensation-3}(b), we also calculate free energy with respect to temperature and it shows that the dashed line part is unstable comparing with solid curve. That means the $T_{max}$ is still phase transition temperature $T_c$ in this case. Below the transition temperature $T_c$, the condensation is a monodrome function of temperature. At the transition temperature, the condensation will jump from finite positive value to zero and the massless AdS hairy black hole solution is unstable comparing with hyperbolic AdS-SW black hole. That is to say hyperbolic AdS-SW black hole is favored when $T\geq T_c$. Below the transition temperature $T\leq T_c$, the condensation is a monodrome function of temperature. At the transition temperature, the condensation will jump from finite positive value to zero. The deformation from ${\nu_3\phi ^3\over L^2}$ does not change the type of phase transition induced by ${\nu_4\phi ^4\over L^2}$ qualitatively.

In summary, we introduce three {types of special deformations} like $\phi^3, \phi^4, \phi^6$ {separately} in massless neutral scalar potential in the bulk. We calculate the condensation of dual operator of the scalar with respect to temperature. From the numerical {behaviors} of condensations, there {exist} transition temperatures $T_{c}=T_{max}$  in three deformations. Furthermore, we calculate the free energy of each deformation to confirm the phase transitions. Finally, these phase transitions induced by three kinds of deformations are the same type qualitatively. Therefore, one can naturally expect that there are still same types of phase transitions in those cases with deformation like superposition of these three kinds of deformations. {We will turn to be rigid in section \ref{section7} to check the stability} of these solutions in the low temperature region $T\leq T_{c}$. In section \ref{section7}, one can find that all these massless hyperbolic hairy AdS black hole are not stable. There exist more stabler solutions, which are in-homogenous solutions. Therefore, from high temperature to low temperature, the AdS hyperbolic black hole will transit to in-homogenous AdS hairy black hole.

\subsubsection{Massive Cases}\label{massivecase}
In this subsection, we would like to deform massive scalar potential by adding $\phi^3, \phi^4, \phi^6$ terms. We mainly focus on obtaining condensation of the dual $\Delta=2$ operator $O_2$ with respect to temperature. We will see there exist phase {transitions} in various deformations and how these deformations affect the phase {transitions} order in details.

\begin{figure}[!h]
\begin{center}
\epsfxsize=6.5 cm \epsfysize=5.5 cm \epsfbox{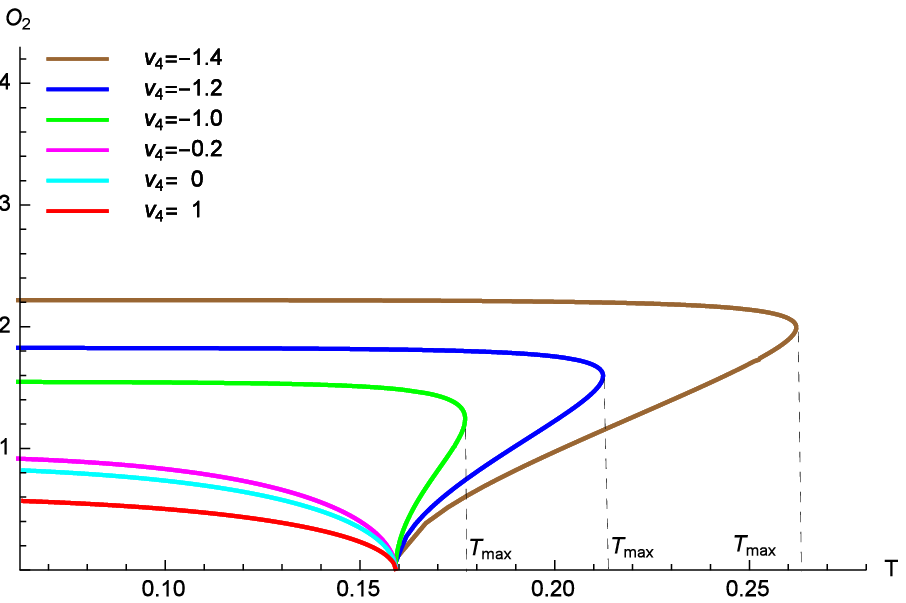}
\hspace*{0.1cm} \epsfxsize=6.5 cm \epsfysize=5.5 cm \epsfbox{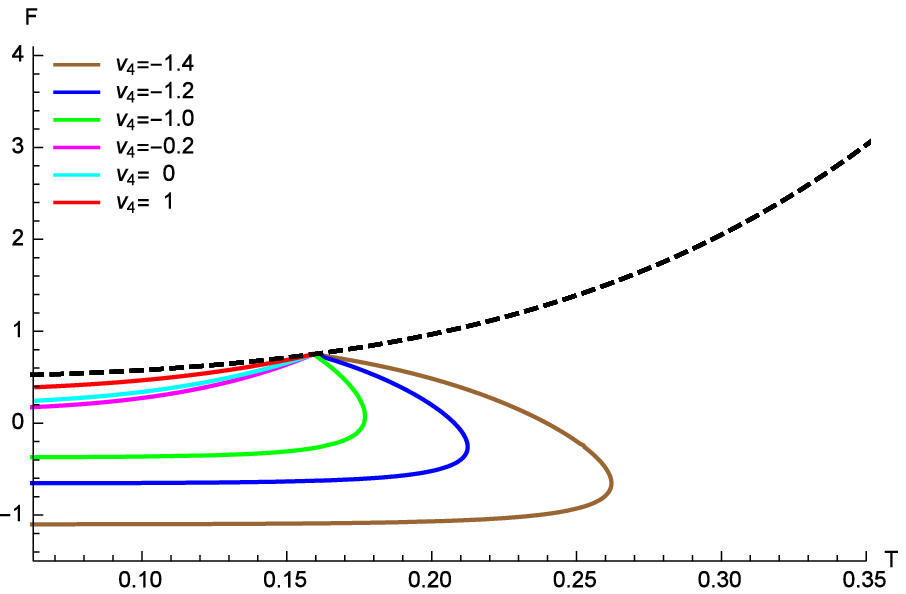}
\hskip 4.5 cm \textbf{( a ) } \hskip 4.5 cm \textbf{( b )}\\
\end{center}
\caption{The condensation is as a function of temperature in massive scalar case with potential $V(\phi )= -{12\over L^2}-{16 \phi ^2\over 3L^2}+{\nu_4\phi ^4\over L^2}$.}\label{condensation-1}
\end{figure}

Firstly, we will turn to study the condensation in massive scalar with potential like $V(\phi )= -{12\over L^2}-{16 \phi ^2\over 3L^2}+{\nu_4\phi ^4\over L^2}$.
In Fig.\ref{condensation-1}(a), we have shown the condensation of dual operator as a function of temperature in several cases. Each case
corresponds to {setting a certain value} of self-interaction coupling constants $\nu_4$. In each case, there is a transition point when the condensation
goes to zero. That means the mass hair AdS hyperbolic black hole is more {stabler} than vanishing condensation solution which is hyperbolic AdS-SW black hole in low temperature region. It implies that there should be a phase transition with increasing temperature in this system. Furthermore, the types of phase transition will
be changed with increasing $\nu_4$, which shows that the ${\nu_4\phi ^4\over L^2}$ deformation will play an important role to determine the transition types. In Fig.\ref{condensation-1}, we increase $\nu_4=-0.2, 0.0, 1.0$ gradually and find that transition temperature is independent on $\nu_4$. Furthermore, there exists a critical value for $\nu_{4c}$ between $\nu_4=-1$ and $\nu_4=-0.2$ . Crossing this critical point, the phase transition order will be changed in $\nu_4< \nu_{4c}$. { As shown in Fig.\ref{condensation-1}, the transition \footnote{Such phase transitions are similar to holographic P-wave Superconductor Phase Transition shown in \cite{Cai:2012nm}.} will be first and second order phase transition in $\nu_4< \nu_{4c}$ and $\nu_4\geq\nu_{4c}$ respectively.} In fig. \ref{condensation-1}(b), the free energy will increase with temperature. All colored curves will converge to one point which corresponds to transition temperature in $\nu_4> \nu_{4c}$. The transition temperature is the same as transition temperature given by fig. \ref{condensation-1}(a). The black dashed line in Fig.~\ref{condensation-1}(b) corresponds to free energy in hyperbolic AdS-SW black hole. In Fig.~\ref{condensation-1}(b), the dominate phase should be hyperbolic AdS-SW black hole above the transition temperature. The free energy can continuously converge to the transition point in Fig.~\ref{condensation-1}(b) with $\nu_4> \nu_{4c}$. But free energy will jump to the transition point with $\nu_4< \nu_{4c}$. That is also means the order of phase transition should change suddenly and the transition temperature will be $T_{max}$, for example, curves shown in $\nu_4=-1.4, -1.2, -1.0$. This phenomenon is also consistent with a condensation jump from finite value to vanishing in Fig.~\ref{condensation-1}(a).

\begin{figure}[!h]
\begin{center}
\epsfxsize=6.5 cm \epsfysize=5.5 cm \epsfbox{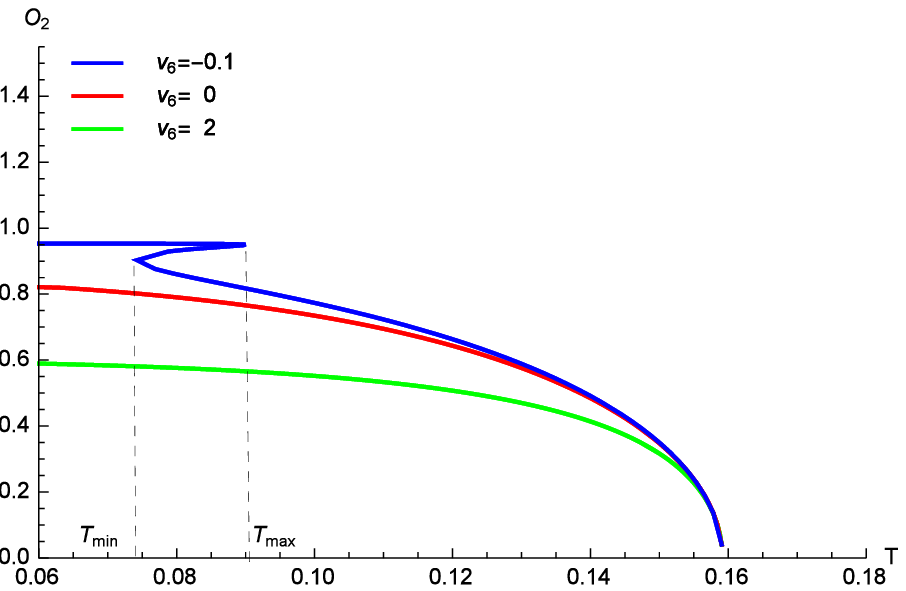}
\hspace*{0.1cm} \epsfxsize=6.5 cm \epsfysize=5.5 cm \epsfbox{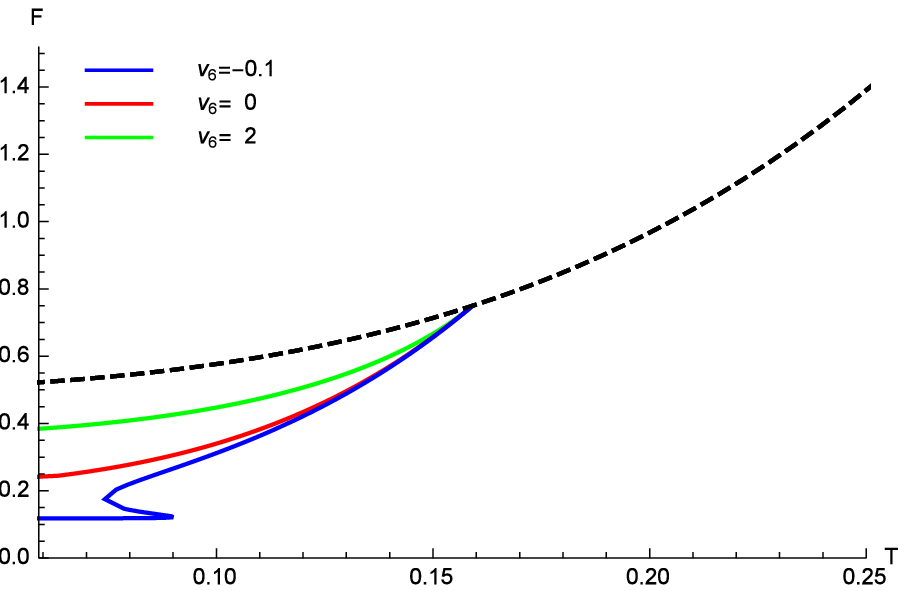}
\hskip 4.5 cm \textbf{( a ) } \hskip 4.5 cm \textbf{( b )}\\
\end{center}
\caption{The condensation $O$ is as a function of temperature $T$ in massive scalar case with potential $V(\phi )= -{12\over L^2}-{16 \phi ^2\over 3L^2}+{\nu_6\phi^6\over L^2}$.}\label{condensation-2}
\end{figure}

Now we will turn to study the condensation in massive scalar with potential like $V(\phi )= -{12\over L^2}-{16 \phi ^2\over 3L^2}+{\nu_6\phi^6\over L^2}$.
We introduce ${\nu_6\phi^6\over L^2}$ deformation and to see what will happen for phase transition. In Fig.~\ref{condensation-2}(a), one can see the condensation with
respect to temperature with choosing different values of coupling constant $\nu_6$. With increasing $\nu_6= 0.0, 2.0$, the condensation will monotonically decrease from
positive finite value to vanishing. In $\nu_6<0.0$ region, the condensation is multiple valued function of temperature as shown in Fig.~\ref{condensation-2}(a) and there is a local maximal temperature $T_{max}$ and minimal temperature $T_{min}$ in each curve. For $\nu_6= -0.1$, the condensation will decrease from $T=0$ to $T=T_{min}$ and it will jump to less finite positive value at $T_{min}$. From $T_{min}<T<T_{max}$, the condensation will become multivalued function of temperature. For $T\geq T_{max}$, the condensation will decrease to zero continuously in Fig.~\ref{condensation-2}(a). In Fig.~\ref{condensation-2}(b), we have shown various free energy with respect to temperature with gradually changing the $\nu_6$. We also find that free energy with $\nu_6= 0, 2$ is monotonically increasing with temperature. They always continuously converge to the transition point $T_c$. The transition point is defined by vanishing of condensation. But in cases with $\nu_6= -0.1$, the free energy is multiple valued function of temperature. For these cases, there are minimal temperatures $T_{min}$ and local maximal temperature $T_{max}$. For $T>T_{c}$, hyperbolic AdS-SW black hole should be stable and there is no massive scalar hair black hole solution. In $T_{max}<T<T_{c}$ and $0< T< T_{min}$, massive scalar hair black hole is more {stabler} than hyperbolic AdS-SW black hole. In $T_{min}<T<T_{max}$, the condensation of dual operator is a multiple valued function and the stable solution is marked by solid curve in Fig.~\ref{condensation-2}(a)(b) in terms of comparing free energy. There is critical value $\nu_{6c}$ such that $T_{min}=T_{max}$. Therefore, there are two types of phase transitions for $\nu_6=-0.5$. The first one happens at $T_{max}$ and the condensation is not continuous function of temperature at $T_{max}$ with $\nu_6<\nu_{6c}$. The other one happens at $T_c$ and condensation goes to zero with $\nu_6>\nu_{6c}$.

\begin{figure}[!h]
\begin{center}
\epsfxsize=6.5 cm \epsfysize=5.5 cm \epsfbox{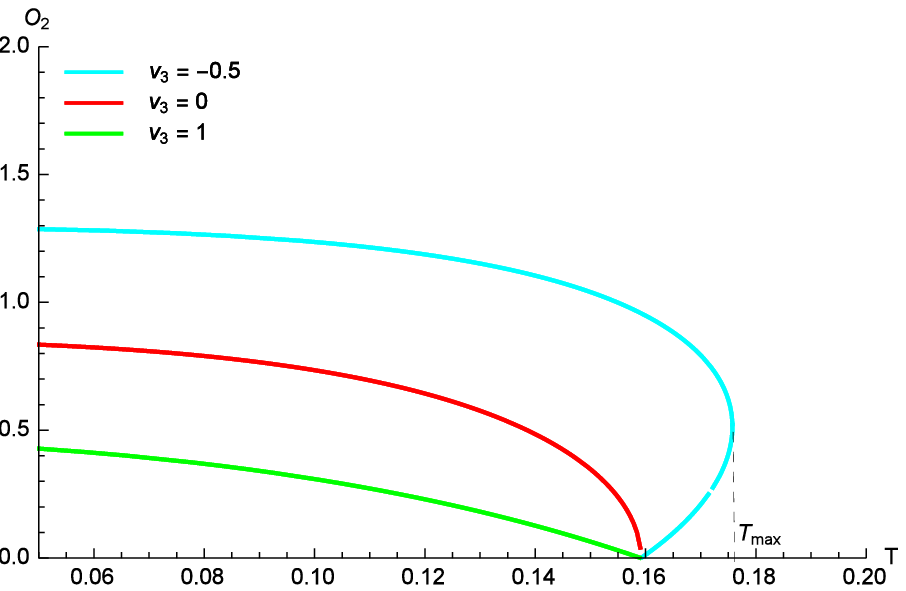}
\hspace*{0.1cm} \epsfxsize=6.5 cm \epsfysize=5.5 cm \epsfbox{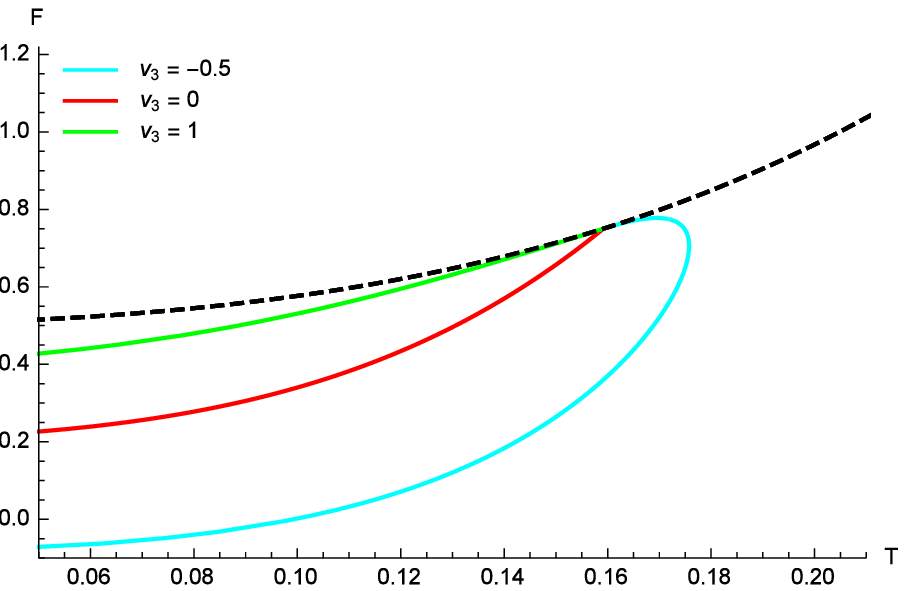}
\hskip 4.5 cm \textbf{( a ) } \hskip 4.5 cm \textbf{( b )}\\
\end{center}
\caption{The condensation $O$ is as a function of temperature $T$ in massive scalar case with potential $V(\phi )= -{12\over L^2}-{16 \phi ^2\over 3L^2}+{\nu_3 \phi^3\over L^2}$.}\label{condensation-4}
\end{figure}

In the third case, we will focus on the condensation with potential $V(\phi )= -{12\over L^2}-{16 \phi ^2\over 3L^2}+{\nu_3 \phi^3\over L^2}$. In Fig.~\ref{condensation-4}(a), we can find that the condensation will decrease from positive finite value to vanishing in $\nu_3>\nu_{3c}$ region. In our setup, $\nu_{3c}=0$. In $\nu_3<0$ region, the condensation will be multiple valued function of temperature. This case is {the similar as} first massive case. In this region, the transition order will be change. {As shown in Fig.\ref{condensation-4}, the transition \footnote{Such phase transitions have also been observed in holographic P-wave Superconductor Phase Transition \cite{Cai:2012nm}. } will be first and second order phase transition in $\nu_3< \nu_{3c}$ and $\nu_3\geq\nu_{3c}$ respectively.} Because the condensation can not continuously decrease to zero at transition temperature and it will suddenly jump from positive finite value to zero. In \ref{condensation-4}(b), we have shown the free energy as function of temperature. In $\nu_3>0$ region, free energy is monotonically increasing with temperature, while free energy is multiple valued function of temperature in $\nu_3<0$. That means the free energy can not converge to the transition temperature continuously, while massive AdS hairy black hole will jump to hyperbolic AdS-SW black hole at transition temperature. Roughly speaking, the phase transitions induced by ${\nu_3 \phi^3\over L^2}$ is {the similar as} ones induced by ${\nu_4\phi^4\over L^2}$.

\begin{figure}[!h]
\begin{center}
\epsfxsize=6.5 cm \epsfysize=5.5 cm \epsfbox{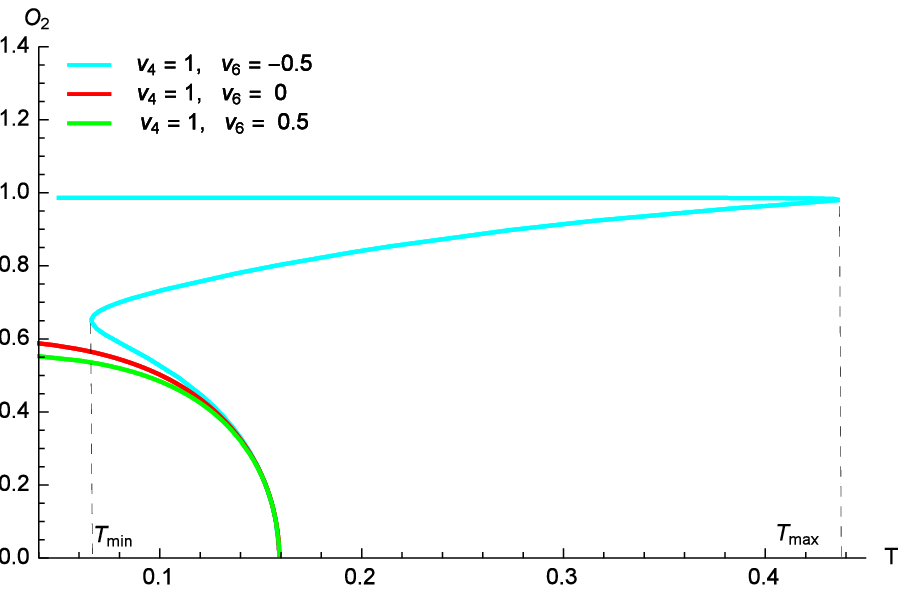}
\hspace*{0.1cm} \epsfxsize=6.5 cm \epsfysize=5.5 cm \epsfbox{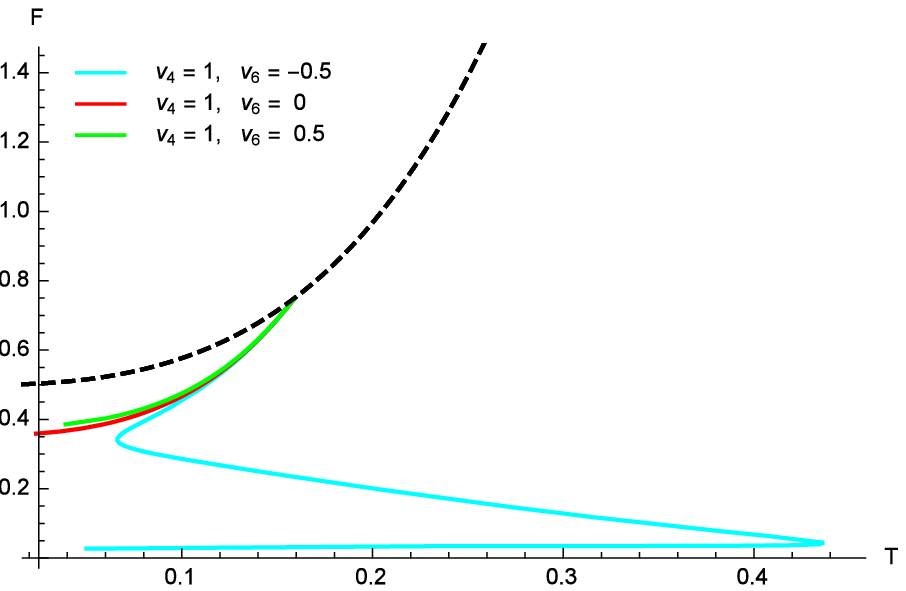}
\hskip 4.5 cm \textbf{( a ) } \hskip 4.5 cm \textbf{( b )}\\
\end{center}
\caption{The condensation $O$ is as a function of temperature $T$ in massive scalar case with potential $V(\phi )= -{12\over L^2}-{16 \phi ^2\over 3L^2}+{\nu_4 \phi^4\over L^2}+{\nu_6\phi ^6\over L^2}$.}\label{condensation-V4-1-V6}
\end{figure}

Finally, we would like to focus on the condensation in massive scalar with potential $V(\phi )= -{12\over L^2}-{16 \phi ^2\over 3L^2}+{\nu_4 \phi^4\over L^2}+{\nu_6\phi ^6\over L^2}$. The main motivation to study this case is that we expect to find competitive mechanism between ${\nu_4\phi ^4\over L^2}$ deformation and $ {\nu_6 \phi^6\over L^2}$ deformation. In Fig.~\ref{condensation-V4-1-V6}(a), one can vary $\nu_6$ with fixing $\nu_4=1.0$ to see that the condensation will monotonically decrease to zero from low temperature to high temperature for $\nu_6>\nu_{6c}$. In $\nu_4=1.0$ case, {$\nu_{6c}=-0.1$} such that $T_{min}=T_{max}$. For fixing $\nu_4$, one can tune $\nu_6=\nu_{6c}$ to be a solution in which $T_{max}$ will coincide with $T_{min}$. One can vary $\nu_4$ to find corresponding $\nu_{6c}$. While in Fig.~\ref{condensation-V4-1-V6}(b), we confirm that the hyperbolic AdS hairy black hole solution is much stable than hyperbolic AdS-SW in $T_{max}<T<T_{c}$ and $0<T<T_{min}$ with $\nu_6<\nu_{6c}, \nu_4=1$. Where $T_c$ is defined by the point where the condensation is vanishing in Fig.~\ref{condensation-V4-1-V6}(a) and $T_{min}, T_{max}$ are marked in Fig.~\ref{condensation-V4-1-V6}(a). In $T>T_{c}$, there is no stable hairy black hole solution for $\nu_6>\nu_{6c}$ and hyperbolic AdS-SW solution is stable one. In $\nu_6<\nu_{6c}$, the condensation will become multivalued function of temperature from $T_{min}<T<T_{max}$,. For $\nu_6=-0.5$ example, the stable configuration in $0<T<T_{min}$ is the hyperbolic AdS hairy black hole solution, while in $T>T_{c}>T_{min}$ is hyperbolic AdS-SW black solution. When $T_{min}<T<T_c<T_{max}$ as shown in Fig.~\ref{condensation-V4-1-V6}, there is a phase transition between two hairy AdS black holes and the condensation will jump from positive finite value to less positive finite value. Especially at $T_c$, there is phase transition between hyperbolic AdS hairy black hole and hyperbolic AdS-SW black hole due to condensation goes to vanishing. These numerical studies show that there is competitive mechanism between ${\nu_4\phi ^4\over L^2}$ deformation and $ {\nu_6 \phi^6\over L^2}$ deformation. One can tune $\nu_4, \nu_6$ to see which phase is stable and what type of phase transition happens. One can set $\nu_4=0$ and this numerical result will reproduce one in second massive case.

Here we have introduced 4 kinds of deformations in massive scalar potential and study these deformations effects on stability of hyperbolic AdS hairy black holes case by case. In each case, the condensation as a function of temperature implies that there exist phase transitions in deformed theories. The behavior of condensation and free energy with respect to temperature in $\phi^3$ and $\phi^4$ deformed theories will be {the same as ones} in the massless cases with $\phi^3, \phi^4, \phi^6$ deformations in section\ref{masslessphasetransition}. Comparing with Fig.\ref{w2-l0-m0-v4-1}  Fig.\ref{w2-l0-m0-v4-v6}  Fig.\ref{w2-l0-m0-v3-v4}, it has different {behaviors} with respect to temperature in $\phi^6$ deformed theories shown in Fig.\ref{condensation-2} Fig.\ref{condensation-V4-1-V6}. Such exotic behavior is induced by $\phi^6$ deformation essentially. In massive cases, they imply that the {types} of phase transitions induced by $\phi^6$ are different from that caused by $\phi^3, \phi^4$. Essentially, all these phase transitions mainly originate from the effective mass of scalar below the effective BF bound for the near horizon $AdS_2$. However, it is not enough to confirm the phase transitions by analyzing the condensation of dual operator and free energy. In section \ref{section7}, we will see the hyperbolic AdS hairy black hole solutions will be stable in low temperature region $T<T_c$ when coupling constants $v_3, v_4, v_6$ live in specific regions. We will see details in section \ref{section7}. {Otherwise, when the coupling constants $v_3, v_4, v_6$ go beyond these specific region, these hyperbolic AdS hairy black hole solutions} will not be stable anymore and there exist much more stabler in-homogenous solutions. {Further numerical studies are needed to check these stability of in-homogenous solutions.}


\section{Instability for the Normalizable Mode}\label{section7}

Previous discussions on the stability of different phases are mainly based on thermodynamical analysis with comparing free energy. Comparing free energy between constant solution \footnote{{We also call hyperbolic AdS hairy black hole solutions}} and hyperbolic AdS-SW is not enough to make sure these new hyperbolic AdS hairy black hole solutions are stable or not. {To be rigorous}, in this section we will investigate the instability of these solutions under scalar perturbation $\delta\Phi(t,z,\psi ,\theta ,\varphi)$. The wave function of $\delta\Phi(t,z,\psi ,\theta ,\varphi)$ could be decomposed as

\begin{eqnarray}\label{pmodes}
\delta\Phi(t, z, \psi ,\theta ,\varphi)=e^{\omega t}\delta\phi(z)Y(\psi ,\theta ,\varphi), \nabla^2_{\mathbf{H}_3}Y(\psi ,\theta ,\varphi)=-\lambda Y(\psi ,\theta ,\varphi),
\end{eqnarray}
with $Y$ the eigenfunction of Laplacian in certain manifold $\Sigma$ and $\lambda$ the corresponding eigenvalues. When $\Sigma$ is just the hyperboloid $H_{d-2}$, $\lambda$ has the lower bound $\lambda>\frac{(d-3)^2}{4}$. Here, we will consider the 5D case, so $d=5$ and $\lambda>1$. More generally, when $\Sigma$ is a non-trivial quotient of hyperboloid, then the lower bound of $\lambda$ would be extended to $0$. Thus, below we will only consider $\lambda>0$ and $\omega^2(\lambda=0)$ for simplicity \cite{Belin:2014lea} \cite{Belin:2013dva}.

Under the ansatz Eq.(\ref{pmodes}), the equation of motion for $\delta\phi$ could be derived as follows
\begin{eqnarray}\label{eom-pet}
\delta\phi^{''}+(-\frac{3}{z}+3A_e^{'}+\frac{f^{'}}{f})\delta\phi^{'}+(\frac{3e^{2A_e}}{8z^2 f}V^{''}(\phi)-\frac{\omega^2}{f^2}+\frac{\lambda}{f})\delta\phi=0,
\end{eqnarray}
where $A_e, f, \phi$ are associated with background solutions. In our ansatz Eq.(\ref{pmodes}), the time related part behaves as $e^{\omega t}$. The black hole will be unstable if (\ref{eom-pet}) has a solution with real and positive $\omega^2$ with the
field satisfying specific boundary conditions at infinity and the horizon. Therefore, if there exist solutions with positive $\omega^2$ in certain background solutions, then the background {constant} solutions are unstable. This {instability} is induced by inhomogenous perturbation in boundary special direction. If one can not find such perturbative modes with positive $\omega^2$, then the background solutions are stable at the level of linear perturbation. This is the key criterion to test the stability of these solutions. In principle, one should construct {AdS hairy black holes} at the non-linear level which is considerably more difficult. In this paper, it is sufficient to demonstrate that an instability exists by linear perturbation.

The leading expansion of $\delta\phi$ near the horizon $z=z_h$ could be derived from Eq.(\ref{eom-pet}) as following
\begin{eqnarray}
\delta\phi(z)=\delta p_{h1} (z_h-z)^{\frac{\omega}{4\pi T}}(1+...)+\delta p_{h2}(z_h-z)^{-\frac{\omega}{4\pi T}}(1+...),
\end{eqnarray}
with $\delta p_1, \delta p_2$ the two integral constants of the second order derivative equation Eq.(\ref{eom-pet}). Without loss of generality, we assume $\omega=\sqrt{\omega^2}>0$, then the $\delta p_{h1}$ mode tends to $0$ when $z$ approaches $z_h$, while the $\delta p_{h2}$ mode is divergent near horizon. Thus, the near horizon boundary condition is easy to be set as $\delta\phi(z_h)=0$.

For the UV boundary condition, again, we could calculate the near boundary expansion of $\delta\phi$ from Eq.(\ref{eom-pet}). It depends on the dimension of $\phi$. For $\Delta=2$ as example, the leading expansion is of the form
\begin{eqnarray}
\delta\phi(z)=\delta p^{2}_{01} z^2 \log(z)+...+\delta p^{(2)}_{02}z^2+....
\end{eqnarray}
As in the background solutions, we will require the coefficient of $z^2\log(z)$ to be $0$. For $\Delta=4$, one can obtain the UV boundary condition as
\begin{eqnarray}
\delta\phi(z)=\delta p^{(0)}_{0}+\delta p^{(4)}_{0}z^4+....
\end{eqnarray}

In general, only for certain groups of $(\lambda, \omega^2)$ the solutions of $\delta\phi$ could satisfy both the UV and IR boundary conditions simultaneously. We will try to find such kind of solutions under the background constant solutions solved in previous sections, and to see whether it is stable or not under the linear perturbation.

\subsection{Massless Scalar Cases}
Firstly, we focus on the stability in the massless cases. In terms of previous arguments in last section, {we only study} the sign of $\omega^2(\lambda=0)$ and we can test stability of these solutions solved in previous several sections.

\begin{figure}[!h]
\begin{center}
\epsfxsize=6.5 cm \epsfysize=5.5 cm \epsfbox{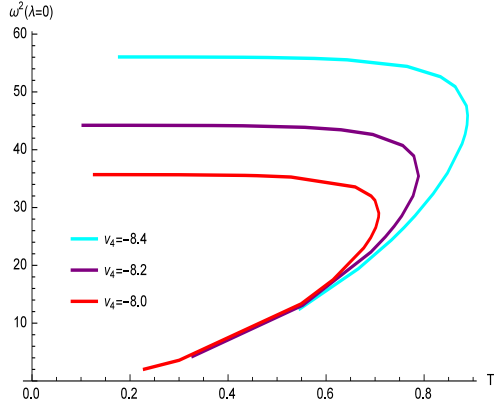}
\end{center}
\caption{$\omega^2(\lambda=0)$ as a function of temperature in massless scalar case with potential $V(\phi )= -{12\over L^2}+{\nu_4 \phi^4\over L^2}$ at the same parameter values as in Fig.\ref{zerocondensation-1}. {Where we have scanned all relevant region of $\nu_4 $ in our setup and we just show most important characteristic qualitative behavior by choosing specific value of $\nu_4$ as examples.} }\label{w2-l0-m0-v4-1}
\end{figure}

\begin{figure}[!h]
\begin{center}
\epsfxsize=6.5 cm \epsfysize=5.5 cm \epsfbox{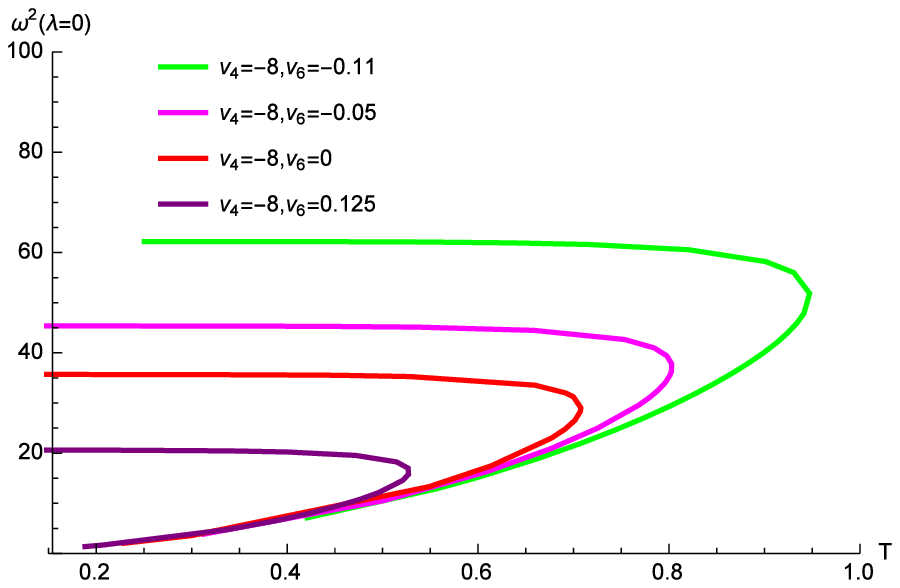}
\end{center}
\caption{$\omega^2(\lambda=0)$ as a function of temperature in massless scalar case with potential $V(\phi )= -{12\over L^2}+{\nu_4 \phi^4\over L^2}+{\nu_6 \phi^6\over L^2}$ at the same parameter values as in Fig.\ref{zerocondensation-2}. {Where we have scanned all relevant region of $\nu_4,\nu_6 $ in our setup and we just show most important characteristic qualitative behavior by choosing specific value of $\nu_4,\nu_6$ as examples.}}\label{w2-l0-m0-v4-v6}
\end{figure}

\begin{figure}[!h]
\begin{center}
\epsfxsize=6.5 cm \epsfysize=5.5 cm \epsfbox{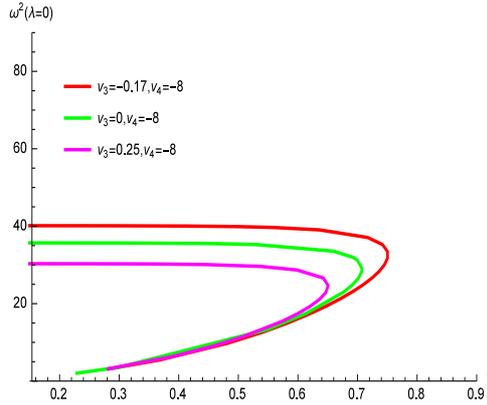}
\end{center}
\caption{$\omega^2(\lambda=0)$ as a function of temperature in massless scalar case with potential $V(\phi )= -{12\over L^2}+{\nu_3 \phi^3\over L^2}+{\nu_4 \phi^4\over L^2}$ at the same parameter values as in Fig.\ref{zerocondensation-3}. {Where we have scanned all relevant region of $\nu_3,\nu_4 $ in our setup and we just show most important characteristic qualitative behavior by choosing specific value of $\nu_3,\nu_4$ as examples.}}\label{w2-l0-m0-v3-v4}
\end{figure}

In Fig.\ref{w2-l0-m0-v4-1} Fig.\ref{w2-l0-m0-v4-v6} Fig.\ref{w2-l0-m0-v3-v4}, we show the $\omega^2(\lambda=0)$ as a function of temperature numerically with turning on the linear perturbation of hyperbolic black hole solution with $V(\phi )= -{12\over L^2}+{\nu_4 \phi^4\over L^2}$ ,$V(\phi )= -{12\over L^2}+{\nu_4 \phi^4\over L^2}+{\nu_6 \phi^6\over L^2}$ and $V(\phi )= -{12\over L^2}+{\nu_3 \phi^3\over L^2}+{\nu_4 \phi^4\over L^2}$ respectively. In all these cases, one can see that $\omega^2(\lambda=0)$ always positive from low to high temperature region. These solutions shown in Fig.\ref{zerocondensation-1} Fig.\ref{zerocondensation-2} Fig.\ref{zerocondensation-3} should be unstable configurations, although these configurations are much more stable than hyperbolic AdS-SW black hole with comparing free energy. One can see that there should exist in-homogenous black hole solutions which break the hyperbolic symmetry. That is also means that hyperbolic AdS-SW black hole will transit to in-homogenous black hole solutions. In-homogenous black hole solutions are hard to be constructed which will be interesting to be studied in the near future.

\subsection{Massive Scalar Cases}
\begin{figure}[!h]
\begin{center}
\epsfxsize=6.5 cm \epsfysize=5.3 cm \epsfbox{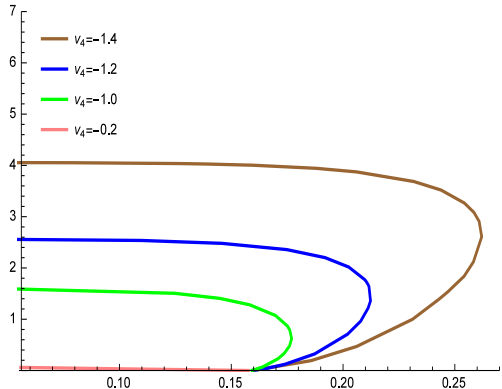}
\end{center}
\caption{$\omega^2(\lambda=0)$ as a function of temperature in massive scalar case with potential $V(\phi )= -{12\over L^2}-{16 \phi ^2\over 3L^2}+{\nu_4 \phi^4\over L^2}$ at the same parameter values as in Fig.\ref{condensation-1}. When $v_4=0,1$, we did not find positive $\omega^2$ at $\lambda=0$. {Where we have scanned all relevant region of $\nu_4 $ in our setup and we just show most important characteristic qualitative behavior by choosing specific value of $\nu_4$ as examples.}}\label{w2-l0-m2-v4}
\end{figure}

\begin{figure}[!h]
\begin{center}
\epsfxsize=6.5 cm \epsfysize=5.3 cm \epsfbox{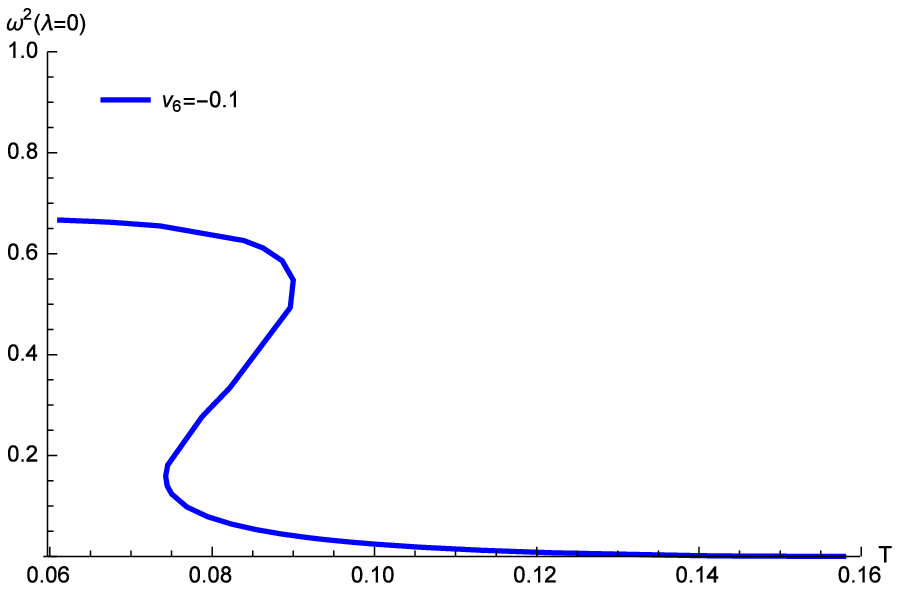}
\end{center}
\caption{$\omega^2(\lambda=0)$ as a function of temperature in massive scalar case with potential $V(\phi )= -{12\over L^2}-{16 \phi ^2\over 3L^2}+{\nu_6 \phi^6\over L^2}$ at the same parameter values as in Fig.\ref{condensation-2}. When $v_6=0,2$, we did not find positive $\omega^2$ at $\lambda=0$. {Where we have scanned all relevant region of $\nu_6 $ in our setup and we just show most important characteristic qualitative behavior by choosing specific value of $\nu_6$ as examples.}}\label{w2-l0-m2-v6}
\end{figure}

\begin{figure}[!h]
\begin{center}
\epsfxsize=6.5 cm \epsfysize=5.3 cm \epsfbox{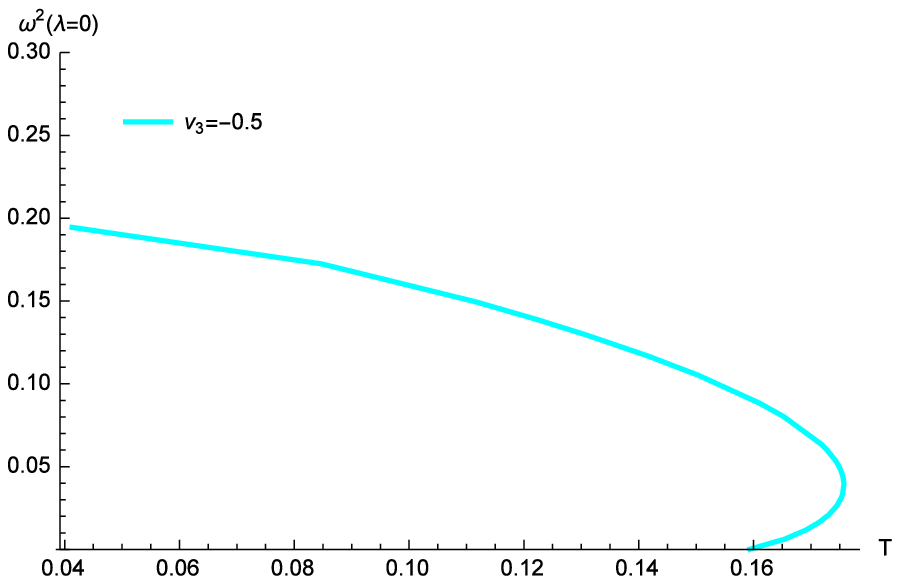}
\end{center}
\caption{$\omega^2(\lambda=0)$ as a function of temperature in massive scalar case with potential $V(\phi )= -{12\over L^2}-{16 \phi ^2\over 3L^2}+{\nu_3 \phi^3\over L^2}$ at the same parameter values as in Fig.\ref{condensation-4}. When $v_3=0,1$, we did not find positive $\omega^2$ at $\lambda=0$. {Where we have scanned all relevant region of $\nu_3$ in our setup and we just show most important characteristic qualitative behavior by choosing specific value of $\nu_3$ as examples.}}\label{w2-l0-m2-v3}
\end{figure}

\begin{figure}[!h]
\begin{center}
\epsfxsize=6.5 cm \epsfysize=5.3 cm \epsfbox{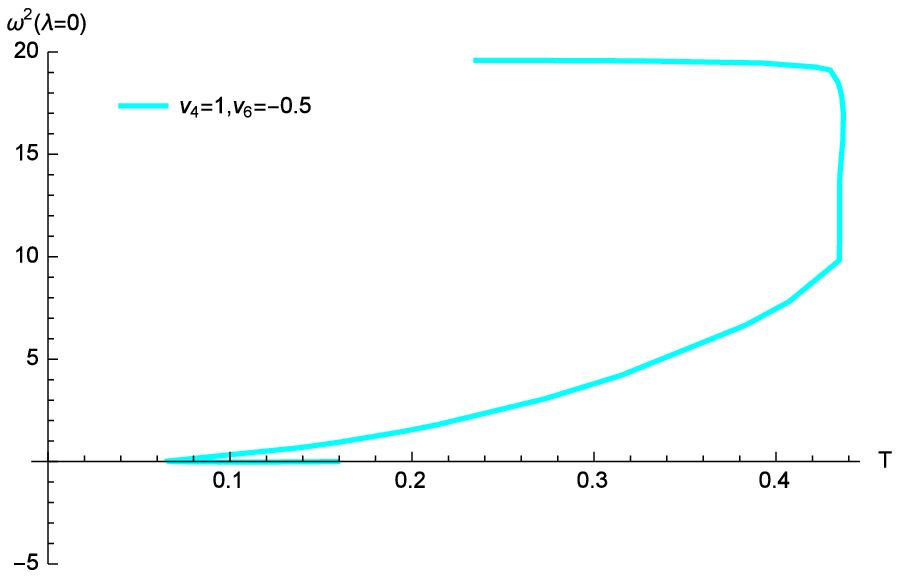}
\end{center}
\caption{$\omega^2(\lambda=0)$ as a function of temperature in massive scalar case with potential $V(\phi )= -{12\over L^2}-{16 \phi ^2\over 3L^2}+{\nu_4 \phi^4\over L^2}+{\nu_6 \phi^6\over L^2}$ at the same parameter values as in Fig.\ref{condensation-V4-1-V6}. When $v_6=-0.1,0,0.5$, we did not find positive $\omega^2$ at $\lambda=0$. {Where we have scanned all relevant region of $\nu_4,\nu_6$ in our setup and we just show most important characteristic qualitative behavior by choosing specific value of $\nu_4,\nu_6$ as examples.}}\label{w2-l0-m2-v4-v6}
\end{figure}

In this subsection, we turn to focus on the stability of new hyperbolic black hole solutions with massive scalar potentials. Here we have studied four cases which are shown in Fig.\ref{w2-l0-m2-v4} Fig.\ref{w2-l0-m2-v6} Fig.\ref{w2-l0-m2-v3} Fig.\ref{w2-l0-m2-v4-v6}. Here we summarize final results in the following.

In Fig.\ref{w2-l0-m2-v4} Fig.\ref{w2-l0-m2-v6} Fig.\ref{w2-l0-m2-v3} Fig.\ref{w2-l0-m2-v4-v6}, we show the $\omega^2(\lambda=0)$ as a function of temperature numerically with turning on the linear perturbation of hyperbolic black hole solution with $V(\phi )= -{12\over L^2}-{16 \phi ^2\over 3L^2}+{\nu_4 \phi^4\over L^2}$ ,$V(\phi )= -{12\over L^2}-{16 \phi ^2\over 3L^2}+{\nu_6 \phi^6\over L^2}$, $V(\phi )= -{12\over L^2}-{16 \phi ^2\over 3L^2}+{\nu_3 \phi^3\over L^2}$ and $V(\phi )= -{12\over L^2}-{16 \phi ^2\over 3L^2}+{\nu_4 \phi^4\over L^2}+{\nu_6 \phi^6\over L^2}$ respectively. For massive scalar cases, there are something new presented. Up to linear perturbative analysis, some homogenous solutions are still stable.

For example, as shown in Fig.\ref{w2-l0-m2-v4}, we can tune $v_4$ gradually and then we can find a critical value of $v_{4c}=0$. Once $v_4>0$, $\omega^2(\lambda=0)$ are always negative definite function of temperature and which also means $v_4>0$ these solutions found in Fig.\ref{condensation-1} might be stable at level of linear perturbation analysis. That means hyperbolic AdS-SW black hole will transit to in-homogenous solution from high temperature to low temperature when $v_3=0, v_6=0, v_4>0$. For $v_3=0, v_6=0, v_4>0$, hyperbolic AdS-SW black hole will\footnote{Here we confirm the transition in terms of linear perturbative analysis.} transit to homogenous solution as shown Fig.\ref{w2-l0-m2-v6}.

One can also tune the $v_6$ gradually to find the critical value of { $v_6=0$.} When $v_6$ becomes positive, one can not find positive definite $\omega^2(\lambda=0)$ which implies that solutions with positive $v_6$ might be also stable and phase transition might happen in Fig.\ref{condensation-2}. That means hyperbolic AdS-SW black hole will transit to homogenous solution from high temperature to low temperature when $v_3=0, v_4=0, v_6\geq0$. The high temperature solution will transit to in-homogenous solutions for $v_3=0, v_4=0, v_6<0$.

In Fig.\ref{w2-l0-m2-v3}, one can also tune the $v_3$ gradually to find critical value $v_3=0$. From high temperature to low temperature, the hyperbolic AdS-SW black hole will transit to in-homogenous solution will transit to in-homogenous solutions for $v_3<0, v_4=0, v_6=0$. For $v_3\geq 0, v_4=0, v_6=0$, it will transit to homogenous solution constructed in section \ref{massivecase}.

Finally, we consider more complicated situation with potential $V(\phi )= -{12\over L^2}-{16 \phi ^2\over 3L^2}+{\nu_4 \phi^4\over L^2}+{\nu_6 \phi^6\over L^2}$. For simplifying our study, we fix $v_3=0, v_4=1$ and gradually tune $v_6$ to obtain the critical value {$v_6=-0.1$} such that there is no positive $\omega^2(\lambda=0)$ existing in the black hole solution. We expect that the solutions found in Fig.\ref{condensation-V4-1-V6} with positive $v_6$ are stable.  In terms of criterion mentioned above, hyperbolic AdS-SW black hole might transit to homogenous solution from high temperature to low temperature when $v_3=0, v_4=1, v_6\geq0$. For $v_3=0, v_4=1, v_6<0$, it will transit to in-homogenous solution as confirmed in Fig.\ref{w2-l0-m2-v4-v6}. Further, one can also see interesting phenomenon that change of $v_4$ will affect the critical value of $v_6$.

%
\section{Comments on Holographic Spherical R\'{e}nyi Entropy}\label{REE}
In this section, we would like to connect the instability of hyperbolic AdS black hole with holographic R\'{e}nyi Entropy, as we reviewed in the introduction.

\subsection{Spherical R\'{e}nyi Entropy as Thermal Entropy}\label{REE6.2}
Following \cite{Belin:2013dva}, we can compute the R\'{e}nyi entropy from these thermal entropies, via (\ref{finalfor111})
 \bea\label{REEphasetransition}
S_{n}={n\over n-1}{1\over T_{0}}
\left(
\int_{T_{0}/{n}}^{T_{cri}}S^{Eh}_{\text{thermal}}(T)dT
+
\int_{T_{cri}}^{T_{0}}S^{E}_{\text{thermal}}(T)dT
\right),
 \eea
where $S^{Eh}_{\text{thermal}}(T)$ is the entropy of the hairy black hole and
$S^{E}_{\text{thermal}}(T)$ is the entropy of the Einstein black hole.

In terms of the above formulas, the R{\'e}nyi entropy as a function of $n$.
Because the derivative of the thermal entropy with respect to the temperature is discontinuous, the second derivative with respect to $n$ of the R{\'e}nyi entropy is discontinuous. Such kind of discontinuous is closely related to instability of hyperbolic AdS black hole. Such instability has been carefully studied in section 4 and section 5. Therefore, the discontinuous of R{\'e}nyi entropy implies a phase transition in dual field theory by holography. In order to determine the
precise value of $n_c$ at which this transition occurs, one should study numerically
the scalar wave equation within the black hole background as shown in \cite{Belin:2013uta}\cite{Belin:2014mva}. The critical temperature is defined by $T_{cri}={1\over 2n_c\pi R}$. {As we shown in section 4 and section 5, the order of phase transition will be different from the second order phase transition \footnote{The same fist order phase transitions have been observed in P-wave Superconductor Phase Transition \cite{Cai:2012nm}.} presented by \cite{Belin:2013dva} due to the {higher powers of neutral scalar with self-interactions.}}

{The first term in left hand side of Eq.(\ref{REEphasetransition}) can work which highly depends on the wether the CHM mapping can work or not in our the hairy black holes. The CHM mapping is a kind of coordinate transformation which simplifies for a spherical or planar entangling surface in CFTs. It works even when conformal symmetry is {partially broken} as long as you take into account the conformal factor eq.(\ref{ansatz}) correctly. It is also clear in the holographic picture where the CHM mapping is a coordinate transformation from the Poincare to the AdS hyperbolic coordinates. It works not only for CFT, but also for any theories with UV fixed points. Indeed, due to the condensation of the scalar, the stable phase will be hyperbolic AdS hairy black hole. {In our cases, the scalar depends only on the holographic direction $z$ and it just modify the conformal factor $e^{A_e(z)}$ in (\ref{ansatz}). The conformal factor only modify the unitary transformation in eq.(\ref{8.1}). }Therefore, CHM mapping is still meaningful in our cases.}

{However, if the scalar depends on {the boundary coordinates called inhomogenous cases in this paper}, CHM mapping can not be applicable in condensation phase with large value of condensation. But in these cases, nearby the transition point $T \sim T_c$\footnote{$|n- n_c|\sim 0 $} from hyperbolic AdS-SW to hyperbolic AdS hairy black hole, the hyperbolic AdS hairy black hole (condensation phase) can be regarded as hyperbolic AdS-SW with neutral scalar perturbatively. And then the condensation of scalar will gradually be turn on and condensation phase will be stable. Near the transition point, the CHM mapping can be still workable approximately due to the small value of condensation. In this sense, CHM mapping gives us an important insight of phase transition in the dual field theory. When hyperbolic AdS hairy black hole become dominant with decreasing temperature, CHM mapping will break down due to large value of condensation and the spherical entanglement surface in dual field can not be identified with thermal entropy of hyperbolic AdS black hole directly.}

{{To be more precisely, once we take $n\gg n_c$,  CHM mapping does not hold any more due to the presence of in-homogenous operator condensation. When $|n- n_c|\sim 0 $ or condensation is very small, as argued in last paragraph, CHM mapping can be still hold approximately. }In this sense, CHM map spherical entanglement entropy to thermal entropy of hyperbolic AdS black hole perturbatively. When $|n- n_c|$ increases gradually, the CHM mapping will be deformed the mapping by two main aspects. The first is that the dual quantum states in field theory will be excited states because of introducing the operator dual to neutral scalar. That means we have to study the R\'{e}nyi entropy of low excited states by CHM mapping approximately. This is closely related to the first law of entanglement entropy\footnote{The first law of entanglement entropy has been well studied in the holographic literature, for example, \cite{Bhattacharya:2012mi}\cite{Guo:2013aca}\cite{Blanco:2013joa}\cite{deBoer:2015kda}.} once the subsystem is very small. The other one is the shape of entangle surface will be deformed\footnote{The entanglement entropy with deformation of entangle surface has been also studied extensively by \cite{Allais:2014ata}\cite{Lewkowycz:2014jia}\cite{Faulkner:2015csl}\cite{Dong:2016wcf}.} from sphere to other general geometrical shape gradually with increasing condensation or $n \to \infty$, due to dominant phase occupied by inhomogenous scalar. In the future, quantitative studies of the two main aspects are needed to show whether and how CHM mapping works in excited states and in-homogenous configurations. However, in our cases, once we take $|n- n_c|\gg 0 $ large enough, what we have calculated by (\ref{8.2}) should correspond to spherical R\'{e}nyi entropy of states excited by operators which dual to the various homogenous scalars respectively.}

To close this section, we offer a short summary to make our claim to be clear.
For the cases of {Einstein Dilaton} system shown in eq.(\ref{minimal-Einstein-action}) eq.(\ref{ansatz}), if the scalars in hyperbolic AdS hairy black hole only depend on holographic coordinate $z$ (called homogenous solutions) and CHM transformation only change the scale transformation by global conformal factor $e^{A_s(z)}$ in (2.3) which preserve the asymptotic AdS boundary conditions, we can make use of CHM mapping to compute R\'{e}nyi entropies. Otherwise, for example, in-homogenous scalar will break CHM transformation and one can not absorb conformal transformation for scalar into global conformal factor anymore. The valid of CHM transformation highly depends on whether we can find CHM transformation up to global conformal factor which correspond to $U$ unitary transformation in eq.(8.1).

\section{Conclusions and Discussions}
{In this paper, we have constructed several new hyperbolic asymptotic AdS gravity solutions in Einstein Dilaton system numerically. Motivated by studying ERE with spherical entangling surface in deformed CFTs by CHM mapping, we work out the hyperbolic hairy AdS with series powers of neutral scalar in potential. In this paper, we focus on potential with $\phi^3, \phi^4, \phi^6$.} Especially, we focus on two kinds of special scalar potentials. The one is massless scalar with higher powers of scalar self interaction and the other is that we choose square of the scalar mass to be $-2$. In terms of AdS/CFT, the first kind of scalar {could correspond} to {dimension 4 gluon sector} in gauge field theory side and the other scalar is dual to {dimension 2 gluon operator} in field theory side. In general, to calculate the ERE with complicated entanglement surface is very hard. For spherical entanglement surface, one can make use of proposals \cite{Casini:2011kv}\cite{Myers:2010xs}\cite{Myers:2010tj} to relate the ERE to the thermal entropy in {hyperbolic AdS black hole. We have shown the configuration of these new hyperbolic AdS solutions and also extract the condensation of operators which are dual to massless and massive scalars respectively. Through studying condensation with respect to temperature, we find that there exist phase transitions.} We list the well defined boundary energy momentum tensor by introducing proper boundary counter terms in each solution. With these counter terms, the finite free energy can be achieved. We compare free energy between the new hyperbolic AdS solutions and hyperbolic AdS-SW solution to check the stability of these solutions. {To be more rigid, we turn on inhomogenous perturbation on these new hyperbolic AdS black holes to {check} the stability. We tune the potential parameters to figure out the stable region of potential parameters for these solutions,} for example, the coefficients of the cubic, quartic and sextic scalar interactions $v_3, v_4, v_6$. For massless scalar cases, we can not find stable homogenous solutions with turning on $\phi^3,\phi^4,\phi^6$ in scalar potentials respectively. Therefore, we can not safely say phase transition shown in Fig.\ref{zerocondensation-1} Fig.\ref{zerocondensation-2} Fig.\ref{zerocondensation-3} really happens. {There must exist stable inhomogenous solutions.} That means hyperbolic AdS-SW black hole will transit to inhomogenous solutions in massless cases from high temperature to low temperature. For massive scalar cases with positive potential parameters $v_3, v_4, v_6$ respectively, $\phi^3,\phi^4$ will induce similar phase transition qualitatively shown in Fig.\ref{condensation-1} Fig.\ref{condensation-4}, {while $\phi^6$ term in scalar potential will induce different kinds of phase transition in Fig.\ref{condensation-2}.} If one turns on superposition of $\phi^3,\phi^4$ and $\phi^6$ in scalar potential, there exists competitive mechanism between phase transitions induced by $\phi^3,\phi^4$ and $\phi^6$ in Fig.\ref{condensation-V4-1-V6}. To be rigid, when we choose negative potential parameters $v_3, v_4, v_6$ respectively, our studies show that all these hyperbolic hairy AdS black hole solutions are not stable ones anymore. {With negative potential parameters $v_3, v_4, v_6$ separately, there may exist stable inhomogenous solutions which are much more stabler than hyperbolic AdS-SW black hole solutions.} In these cases, hyperbolic AdS-SW black hole solutions will transit to inhomogenous solutions from high temperature to low temperature. {Once $v_4, v_6$ are turn on simultaneously}, the critical values of $v_4, v_6$ will be changed accordingly respectively. {From this phenomenon, one can expect that there exist competition mechanism to determine the critical value of potential parameters $v_3, v_4, v_6$. Once we know the phase structures of various black hole solutions, we make use of \cite{Casini:2011kv}\cite{Myers:2010xs}\cite{Myers:2010tj} to comment on the phase structure in the dual field theory in terms of  spherical entanglement entropy. In this sense, the stability of these black hole solutions is closely related to the spherical ERE in holographic dual CFTs and it gives some insight of phase transitions in dual field theory.}

In this paper, we focus on massless and massive scalar cases with higher powers of self interaction in potentials. In general, such kinds of deformations will lead to various types of phase transitions which are highly sensitive to the operators chosen and types of deformations. ERE can be also regarded as an order parameter to give some insight on phase transitions in dual field theories. Finally, we analysis how CHM can work in our cases. For our gravity setup, we offer a circumscribed criterion to judge whether CHM can work or not. For generic setup, we offer an idea to use CHM mapping to study R\'{e}nyi entropy in perturbative sense in section \ref{REE6.2} and it is still open question.

\section*{Acknowledgements}
We are grateful to Janet Hung, Li Li, S.~Matsuura, Tatsuma Nishioka, Tadashi Takayanagi and Stefan Theisen for very useful conversations and correspondences. Z.F. and D.L. thank Yue-Liang Wu for his supports. S.H. thanks Ronggen Cai, Tadashi Takayanagi, Stefan Theisen for their encouragements and supports. S.H. is supported by Max-Planck fellowship and by the National Natural Science Foundation of China (No.11305235). D.L. is supported by China Postdoctoral Science Foundation.
\vspace*{5mm}

\appendix
\section{Asymptotic AdS Solutions}\label{Appendix8}
In this appendix, we would like to show some details how to obtain these numerical solution.
{Basing} on the set up in section \ref{gravity-setup}, we pay attention to how to solve the whole system in the UV region $z\sim 0$ in this section. Near the UV region, we can use series expansion to find the solution of unknown {functions} in metric ansatz (\ref{ansatz}). These expansions will be helpful to the later numerically computation to show the full numerical solutions.

\subsection{Massless Scalar Cases}\label{app-a1}
In this section, we will try to find the UV expansion of gravity solution with massless scalar with potential like
\bea
V=\frac{1}{L^2}\Big(-12+v_3 \phi ^3+v_4 \phi ^4+v_6 \phi ^6\Big)
\eea In this potential, we set the mass of the scalar to be zero and call this case by massless scalar case for convenience in this paper.

Firstly, the UV behavior of the black hole should be asymptotical
AdS and there is a horizon parameterized
by $z_h$ in the IR region. We find an algorithm to get the numerical solution
consistently. Roughly speaking, we try to  expand in
power series all unknown functions as positive powers of $z$. The UV solution can be expressed by following form
\bea\label{sol1phi}
\phi(z)&=&p_4 z^4+\frac{2 p_4 z^6}{3 L^2}+\frac{z^8 \left(p_4-f_4 L^4 p_4\right)}{2 L^4}-\frac{2 p_4 z^{10} \left(2 f_4 L^4-1\right)}{5 L^6}\nonumber\\&+&\frac{z^{12} \left(1728 f_4^2 L^8 p_4-5184 f_4 L^4 p_4+81 L^8 p_4^3 v_4+512 L^8 p_4^3+1728 p_4\right)}{5184 L^8}\nonumber\\&+&\frac{z^{14}}{997920 L^{10}} \Big(855360 f_4^2 L^8 p_4-1140480 f_4 L^4 p_4+34749 L^8 p_4^3 v_4\nonumber\\&+&217600 L^8 p_4^3+285120 p_4\Big)+O(z^{16})
\eea
\bea\label{sol1Ae}
A_e(z)&=&\frac{1}{81} (-8) p_4^2 z^8-\frac{64 p_4^2 z^{10}}{495 L^2}+\frac{64 p_4^2 z^{14} \left(3 f_4 L^4-2\right)}{945 L^6}+\frac{16 z^{12} \left(2 f_4 L^4 p_4^2-3 p_4^2\right)}{351 L^4}\nonumber\\&+&\frac{z^{16} \left(-69984 f_4^2 L^8 p_4^2+279936 f_4 L^4 p_4^2-2187 L^8 p_4^4 v_4-11776 L^8 p_4^4-116640 p_4^2\right)}{892296 L^8}\nonumber\\&-&\frac{2 z^{18} \left(285120 f_4^2 L^8 p_4^2-475200 f_4 L^4 p_4^2+8019 L^8 p_4^4 v_4+43520 L^8 p_4^4+142560 p_4^2\right)}{2285415 L^{10}}+O(z^{20})\nonumber\\
\eea
\bea\label{sol1f}
f(z)&=&1-\frac{z^2}{L^2}+\frac{32 p_4^2 z^{14} \left(74 f_4 L^4-33\right)}{15015 L^6}+f_4 z^4-\frac{32 p_4^2 z^{10}}{405 L^2}\nonumber\\
&+&\frac{8 z^{12} \left(11 f_4 L^4 p_4^2-9 p_4^2\right)}{891 L^4}-\frac{8 z^{16} \left(15 f_4^2 L^8 p_4^2-42 f_4 L^4 p_4^2+13 p_4^2\right)}{1755 L^8}\nonumber\\&+&\frac{z^{18} \left(-645408 f_4^2 L^8 p_4^2+819072 f_4 L^4 p_4^2-3645 L^8 p_4^4 v_4-48640 L^8 p_4^4-194400 p_4^2\right)}{3903795 L^{10}}+O(z^{20})\nonumber\\
\eea

One can find the black hole solution in the UV region can be
expressed in series of powers of $z$. In principle, one can obtain
more higher powers of $z$ to get the full expression of black hole
background. Unfortunately, we can not obtain closed form of the
black hole solution. The main reason is that we {can} not find simple
recurrence relation among the coefficients of each power of $z$, as explained in \cite{He:2013rsa}. In terms of AdS/CFT dictionary, the massless neutral scalar in the bulk will dual to $\Delta=4$ operator in field theory side. {$p_4$ is the expectation value of dual operator $\langle O_2 \rangle$ with turning off source term in $\phi$ eq.(\ref{sol1phi}).} It
is easy to see that the black hole solution with asymptotical AdS
can be controlled by integral constants $p_4, f_{4}$ in (\ref{sol1phi})(\ref{sol1Ae})(\ref{sol1f}).
$p_4, f_{4}$ are determined by boundary
condition in IR region. Here we choose parameters $p_4, f_{4}$
to show one black hole solution numerically. Here $ p_4, f_{4}$ are
not independent and they are related to the horizon position $z_h$
such that $Q(z_h)=0$. We impose $\phi(z_h)$ to be regular, which could be guaranteed by requiring $Q(z_h)=0$.

\subsection{Massive Scalar Cases}\label{app-a2}
Firstly, we try to figure out asymptotic AdS solution of our setup with potential like
\bea
V=\frac{1}{L^2}\Big(-12-\frac{16 \phi ^2}{3}+v_3 \phi ^3+v_4 \phi ^4+v_6 \phi ^6\Big)
\eea In this potential, we have introduced a mass term of scalar field and we will call this case by massive scalar case.
With above potential, we can find the solution near the UV region analytically. As shown in massless case, the UV
behavior of the black hole should be asymptotical AdS and there is a
horizon in the IR region which is parameterized by $z_h$. The asymptotic solution is following
\bea
\phi(z)&=&p_2 z^2+p_{22} z^2 \log (z)+\frac{1}{64} z^4 \left(18 p_2^2 v_3-36 p_{22} p_2 v_3+27 p_{22}^2 v_3+32 p_{22}\right)+\frac{9}{32} p_{22}^2 v_3 z^4 \log ^2(z)\nonumber\\&+&\frac{9}{16} z^4 \left(p_2 p_{22} v_3-p_{22}^2 v_3\right) \log (z)+O(z^6 \log(z))\nonumber\\
&-&\frac{z^6}{18432000 L^4} \Big(4608000 f_4 L^4 p_2+2304000 f_4 L^4 p_{22}-729000 L^4 p_2^3 v_3^2\nonumber\\&+&1366875 L^4 p_{22}^3 v_3^2-3371625 L^4 p_2 p_{22}^2 v_3^2+2551500 L^4 p_2^2 p_{22} v_3^2-1728000 L^4 p_2^3 v_4\nonumber\\&+&648000 L^4 p_{22}^3 v_4-1944000 L^4 p_2 p_{22}^2 v_4+2592000 L^4 p_2^2 p_{22} v_4-3276800 L^4 p_2^3
\nonumber\\&+&595968 L^4 p_{22}^3-2043904 L^4 p_2 p_{22}^2+2129920 L^4 p_2^2 p_{22}-2592000 L^2 p_2^2 v_3\nonumber\\&-&1620000 L^2 p_{22}^2 v_3+2592000 L^2 p_2 p_{22} v_3-4608000 p_{22}\Big)+O(z^8)\label{massivescalar}
\eea
\bea
A_e(z)&=&\frac{z^6 \left(-74088 p_2^3 v_3+142884 p_{22} p_2^2 v_3-96390 p_{22}^2 p_2 v_3-14013 p_{22}^3 v_3-131712 p_{22} p_2-25088 p_{22}^2\right)}{1555848}\nonumber\\&-&\frac{1}{21} p_{22}^3 v_3 z^6 \log ^3(z)+\frac{1}{98} z^6 \left(9 p_{22}^3 v_3-14 p_2 p_{22}^2 v_3\right) \log ^2(z)\nonumber\\&+&\frac{z^6 \left(-2295 p_{22}^3 v_3+6804 p_2 p_{22}^2 v_3-5292 p_2^2 p_{22} v_3-3136 p_{22}^2\right) \log (z)}{37044}\nonumber\\&+&\frac{\left(-200 p_2^2-20 p_{22} p_2-21 p_{22}^2\right) z^4}{2250}-\frac{4}{45} p_{22}^2 z^4 \log ^2(z)-\frac{2}{225} \left(p_{22}^2+20 p_2 p_{22}\right) z^4 \log (z)+O(z^6)\nonumber\\
\eea
\bea
f(z)&=&1-z^2f_4 z^4-\frac{2 \left(900 p_2^2-660 p_{22} p_2+407 p_{22}^2\right) z^6}{10125}\nonumber\\&-&\frac{1}{45} 8 p_{22}^2 z^6 \log ^2(z)-\frac{8}{675} \left(30 p_2 p_{22}-11 p_{22}^2\right) z^6 \log (z)+O(z^8)
\eea

It is easy to see that the black hole solution with
asymptotical AdS can be controlled by three integral constants
$p_2, p_{22} , f_{4}$.
$p_2, p_{22} , f_{4}$ are determined by boundary
condition in IR region. In this case, $p_2, p_{22} , f_{4}$ are not
independent and they are determined by the black hole horizon $z_h$.
We still impose $\phi(z_h)$ to be regular which is horizon boundary condition.
The temperature is also defined by $T={{f'(z)\over 4\pi}|_{z=z_h}}$. In terms of AdS/CFT dictionary, {$p_{22}$ is source of the expectation value of dual operator $\langle O_2 \rangle=p_2$ with setting vanishing coefficient of $z^2 \log (z)$ in $\phi$ eq.(\ref{massivescalar}).}

\section{Appendix: New Hyperbolic Black Hole Solutions}\label{Appendix9}
In this paper, we focus on the scalar potential with polynomial form of scalar with highest sextic self-interaction. We explore a systematic way to generate fully backreaction gravity solutions and investigate corresponding phase structure. In the following subsections, we will show two examples to demonstrate these configurations of fields.

\subsection{Massless Scalar Cases}
In this subsection, we will show how to solve the gravity background with potential $V(\phi)=\frac{1}{L^2}\left(-12+v_4 \phi^4 \right)$. Here we set the mass term of scalar to be vanishing. In this case, the dual operator $O_1$ is relate to dimension 4 glueball operator. We will take $v_4=8$ as an example to show the numerical process.

As was shown in \ref{app-a1}, the near boundary expansion of the {equations of motion contain} only contains two integral constants $p_4, f_4$. At a first sight, $p_4$ and $f_4$ should be independent on each other, since both of them are integral constants of the non-linear ordinary derivative equations. However, as we mentioned in Sec.\ref{gravity-setup}, the non-linear ordinary equations require a regular condition $Q(z_h)=0$ at horizon $z=z_h$ naturally. This additional IR boundary condition would require the two integral constants extracted in UV region to be dependent on each other. Thus only one integral constant is free, which is related closely to the only relevant physical quantity--temperature.

\begin{figure}[!h]
\begin{center}
\epsfxsize=4.5 cm \epsfysize=4.5 cm \epsfbox{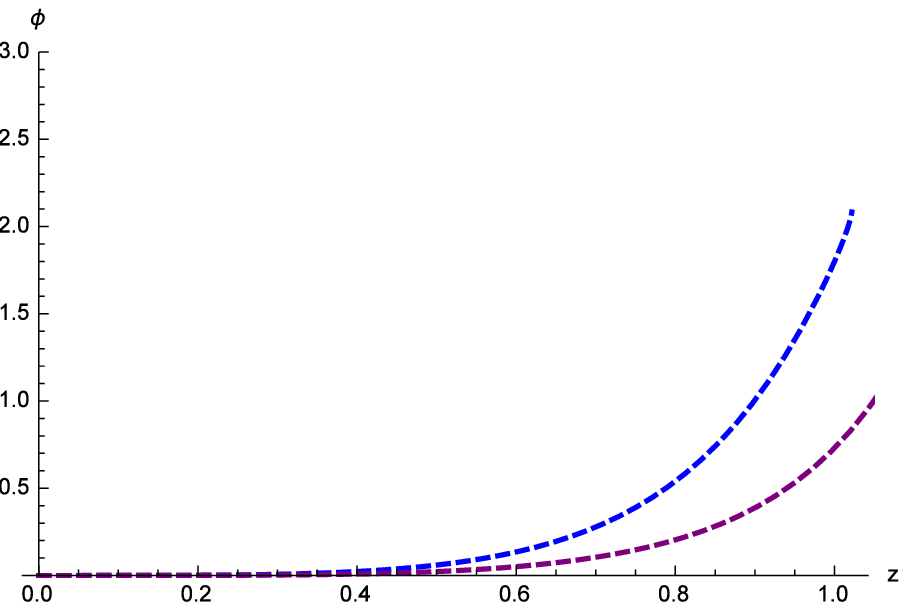}
\hspace*{0.1cm} \epsfxsize=4.5 cm \epsfysize=4.5
cm\epsfbox{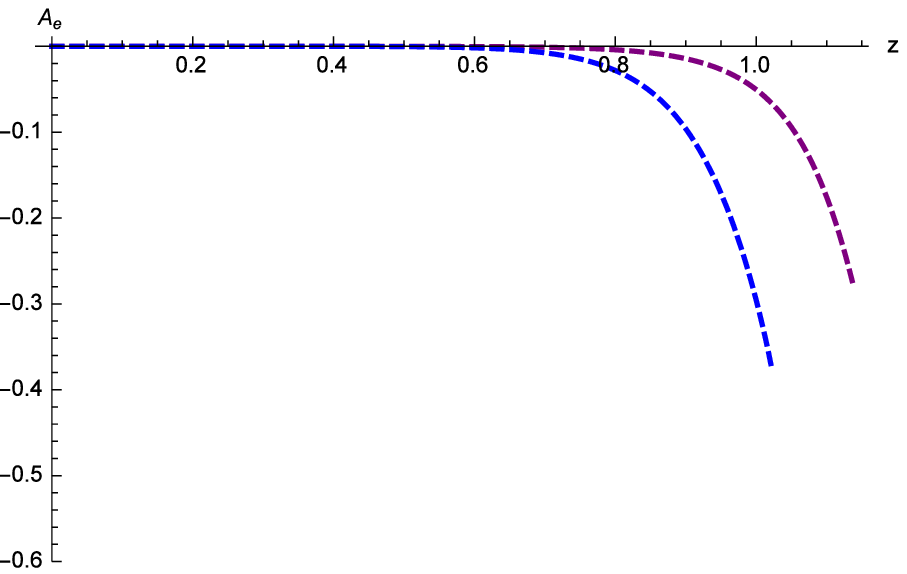}\hspace*{0.1cm} \epsfxsize=4.5 cm \epsfysize=4.5
cm\epsfbox{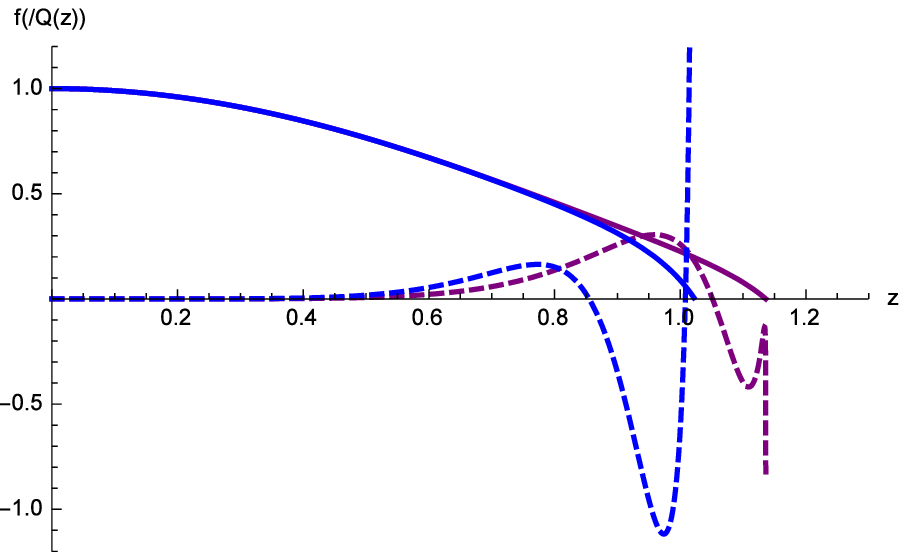}  \vskip -0.05cm \hskip 0.15 cm
\textbf{( a ) } \hskip 4.5 cm \textbf{( b )}\hskip 4.5 cm  \textbf{( c )}\\
\end{center}
\caption{Solutions when $V(\phi )= -{12\over L^2}+{\nu_4\phi ^4\over L^2}$ with $\nu_4=-8, f_4=0.2445$. The blue lines give the results when $p_4=0.3$ while the purple lines are for $p_4=0.8$. In Panel.(a) and Panel.(b), the solutions of $\phi$ and $A_e$ are given. In Panel.(c), the solutions of $f$ is shown in red solid line, while the corresponding $Q(z)$ is shown in blue dashed line(Here, in order to put the two in the same figures, we plot $Q(z)/50$, which is zero at the same $z$ as $Q(z)$).} \label{massless-11}
\end{figure}

\begin{figure}[!h]
\begin{center}
\epsfxsize=4.5 cm \epsfysize=4.5 cm \epsfbox{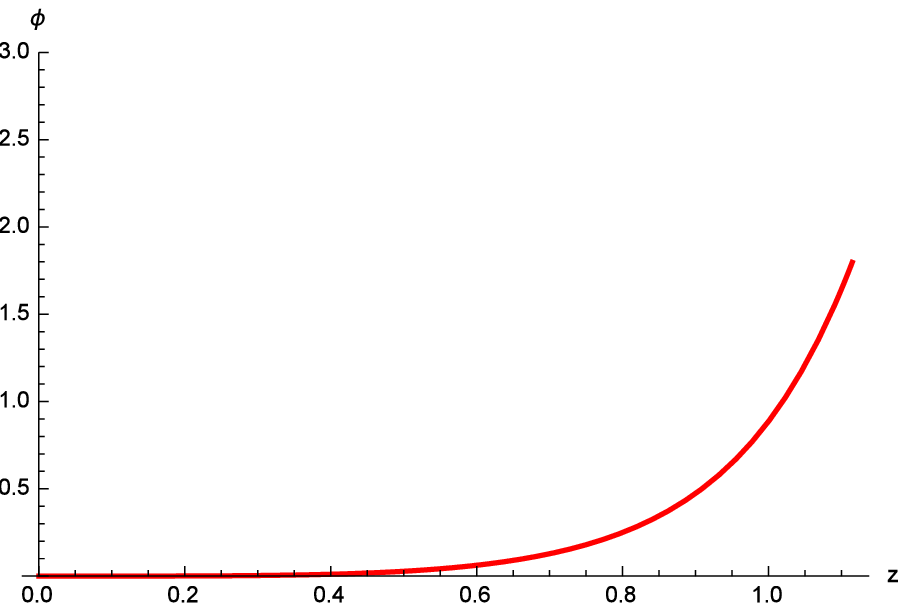}
\hspace*{0.1cm} \epsfxsize=4.5 cm \epsfysize=4.5
cm\epsfbox{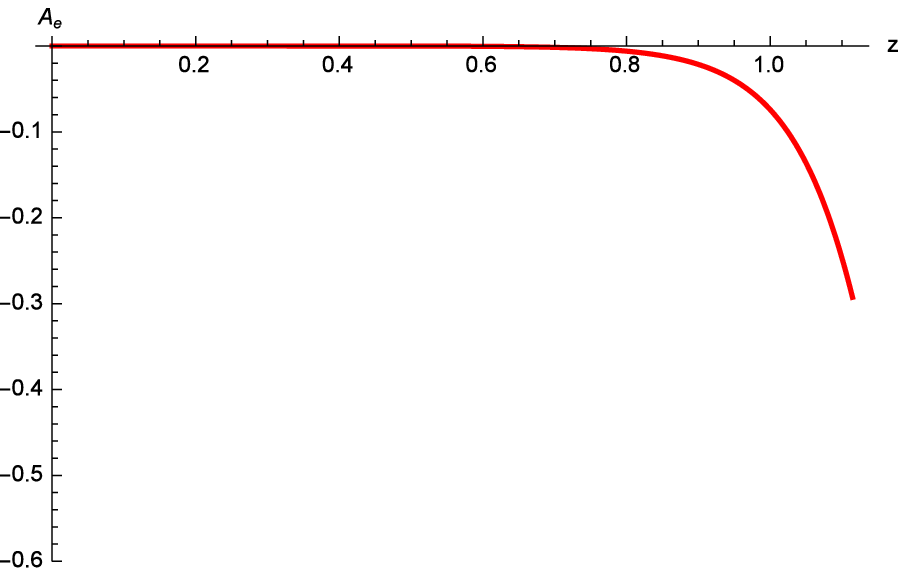}\hspace*{0.1cm} \epsfxsize=4.5 cm \epsfysize=4.5
cm\epsfbox{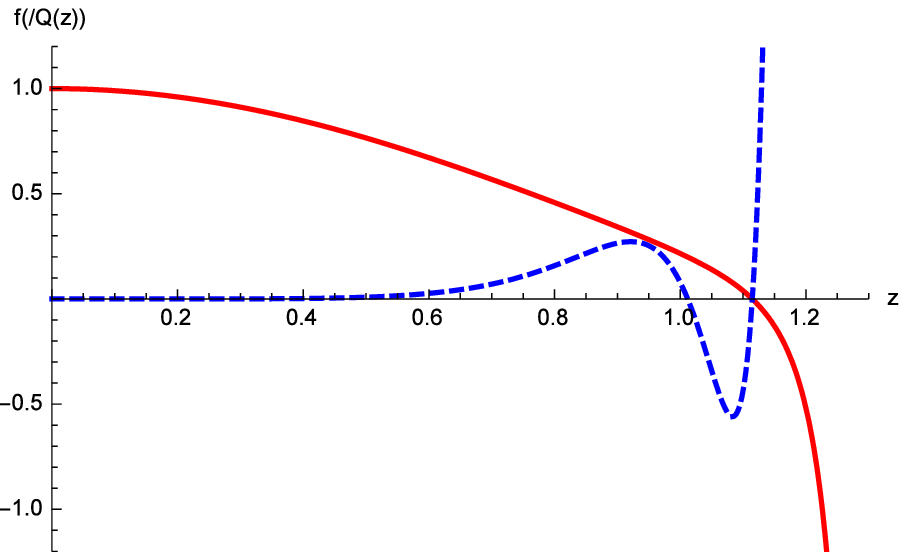}  \vskip -0.05cm \hskip 0.15 cm
\textbf{( a ) } \hskip 4.5 cm \textbf{( b )}\hskip 4.5 cm  \textbf{( c )}\\
\end{center}
\caption{Characteristic solutions when $V(\phi )= -{12\over L^2}+{\nu_4\phi ^4\over L^2}$ with $\nu_4=-8$. To get these solutions, we have taken $f_4=0.2445,p_2=0.36734...$. In Panel.(a) and Panel.(b), the solutions of $\phi$ and $A_e$ are given. In Panel.(c), the solutions of $f$ is shown in red solid line, while the corresponding $Q(z)$ is shown in blue dashed line(Here, in order to put the two in the same figures, we plot $Q(z)/50$, which is zero at the same $z$ as $Q(z)$).} \label{massless-1}
\end{figure}

In the numerical method we used, we try to find the dependence of $p_4$ on $f_4$. As an example, we take $f_4=0.2445$. Now we have the boundary value problem $f_4=0.2445$ and $Q(z_h)=0$. To solve it, we take a test value $p_4=0.3$ and integral the derivative equations from UV to IR with the UV expansion as the initial boundary condition. The results are shown in blue lines in Fig.\ref{massless-11}. From Fig.\ref{massless-11}.(c) we could see that $Q(z)$ blows up when $f(z)$ approaching $0$. This indicates that $p_4=0.3$ is not a regular solution. Therefore, we increase $p_4$ to $p_4=0.8$, and again we see that $Q(z)$ blows up near $f(z)\rightarrow0$. From Fig.\ref{massless-11}.(c), we could see that $Q(z)\rightarrow\pm \infty$ for the two cases, so the proper value of $p_4$ should located at $0.3<p_4<0.8$. Then we take $p_4=\frac{0.3+0.8}{2}$ and solve the solutions. Then if the behavior of $Q(z)$ is like $p_4=0.3$, we have to choose a new value of $p_4$ in the range $0.55<p_4<0.8$, while if it is like $p_4=0.8$, we have to choose in the range $0.3<p_4<0.55$. Repeating the process, finally, we find that when $p_4=0.36734...$, all the quantities could go through the horizon regularly, as shown in Fig.\ref{massless-1}. It is easy to understand that this process could have high accuracy in determine the exact value of $p_4$, since $f(z_h)=0$ in the denominator requires an exact condition $Q(z_h)=0$ to cancel the singularity.


\subsection{Massive Scalar Cases}
\begin{figure}[!h]
\begin{center}
\epsfxsize=4.5 cm \epsfysize=4.5 cm \epsfbox{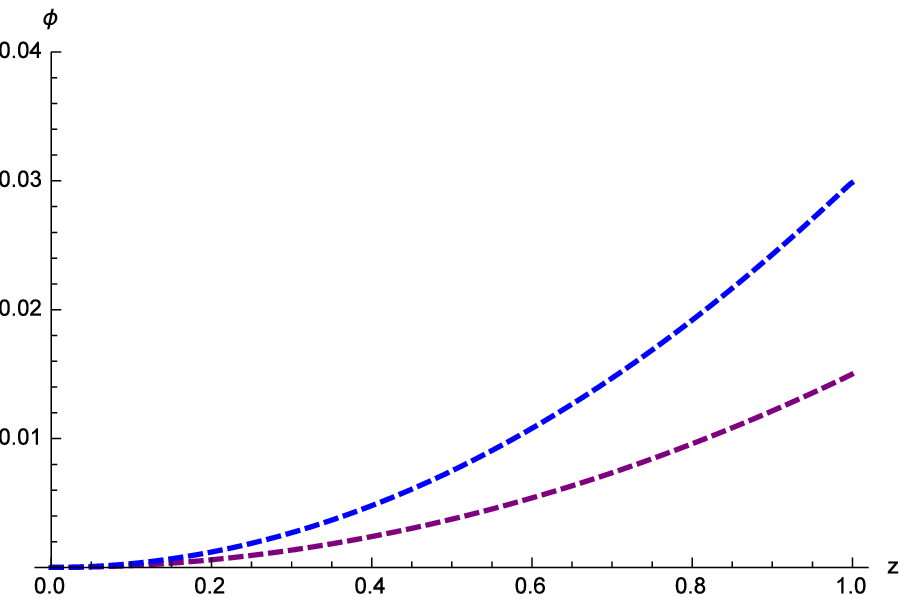}
\hspace*{0.1cm} \epsfxsize=4.5 cm \epsfysize=4.5
cm\epsfbox{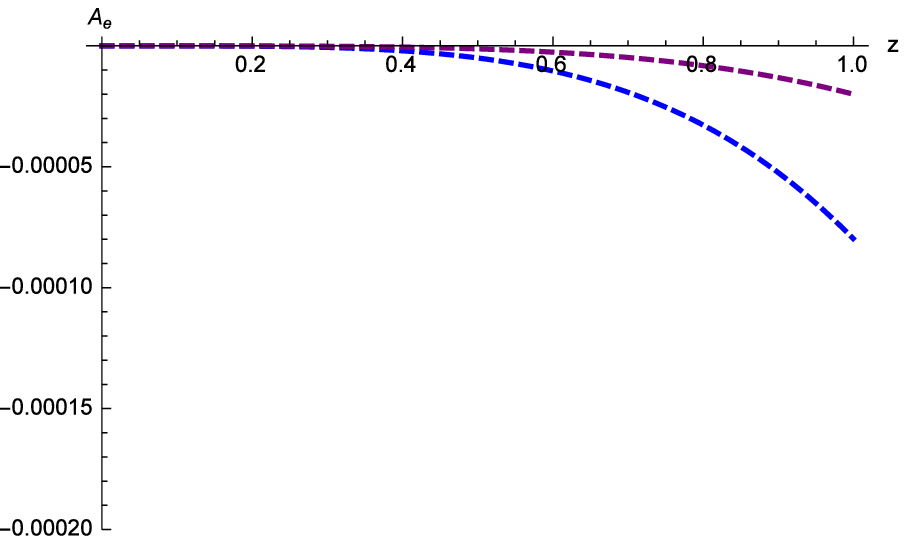}\hspace*{0.1cm} \epsfxsize=4.5 cm \epsfysize=4.5
cm\epsfbox{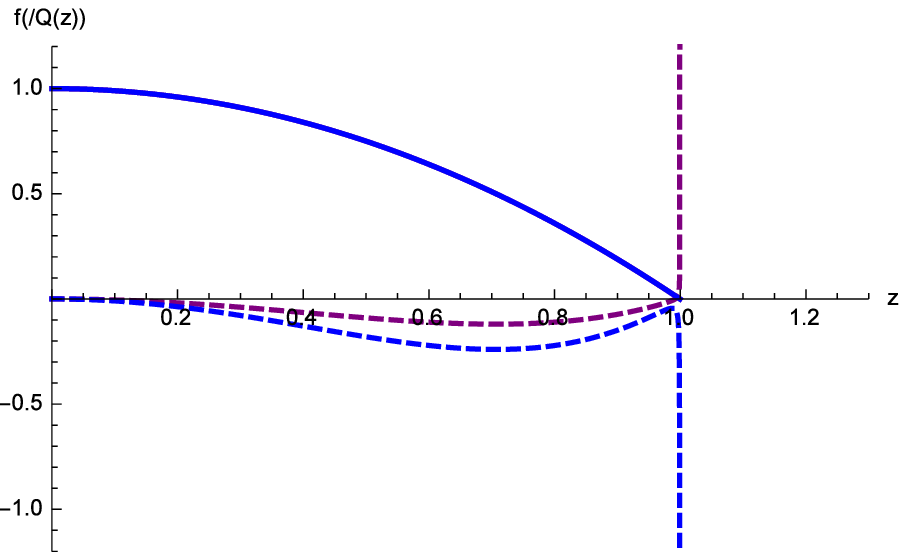}  \vskip -0.05cm \hskip 0.15 cm
\textbf{( a ) } \hskip 4.5 cm \textbf{( b )}\hskip 4.5 cm  \textbf{( c )}\\
\end{center}
\caption{Solutions when $V(\phi )= -{12\over L^2}-{16 \phi ^2\over 3L^2}+{\nu_4\phi ^4\over L^2}$ with $\nu_4=-8, f_4=-0.001$. The blue lines give the results when $p_2=0.03$ while the purple lines are for $p_2=0.015$. In Panel.(a) and Panel.(b), the solutions of $\phi$ and $A_e$ are given. In Panel.(c), the solutions of $f$ is shown in red solid line, while the corresponding $Q(z)$ is shown in blue dashed line.} \label{massive-11}
\end{figure}

\begin{figure}[!h]
\begin{center}
\epsfxsize=4.5 cm \epsfysize=4.5 cm \epsfbox{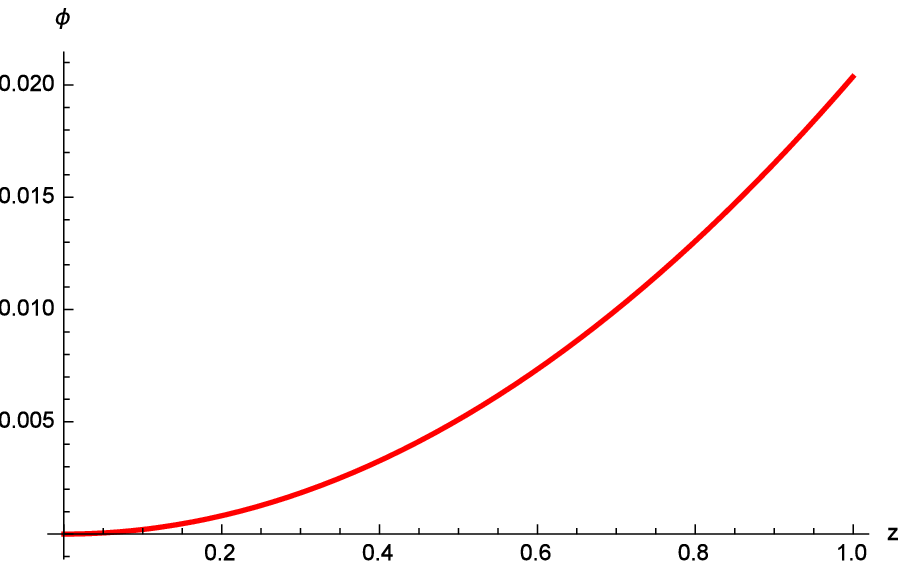}
\hspace*{0.1cm} \epsfxsize=4.5 cm \epsfysize=4.5
cm\epsfbox{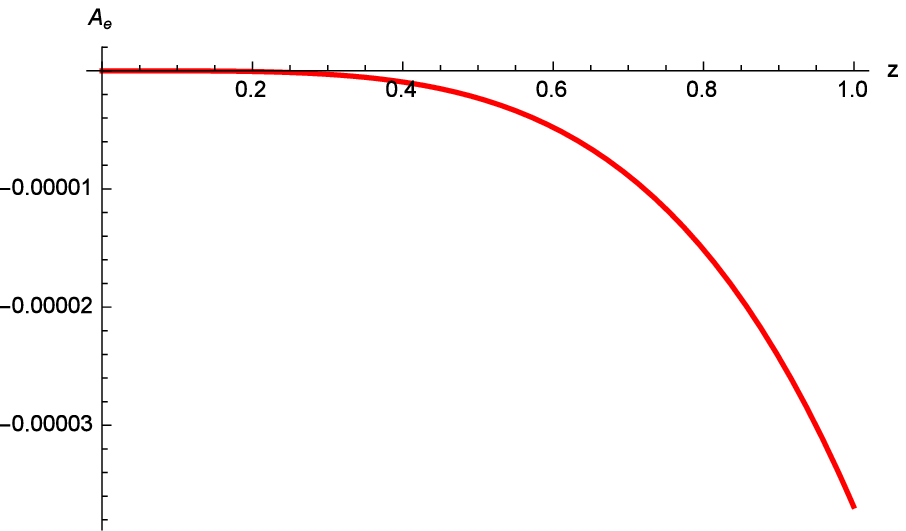}\hspace*{0.1cm} \epsfxsize=4.5 cm \epsfysize=4.5
cm\epsfbox{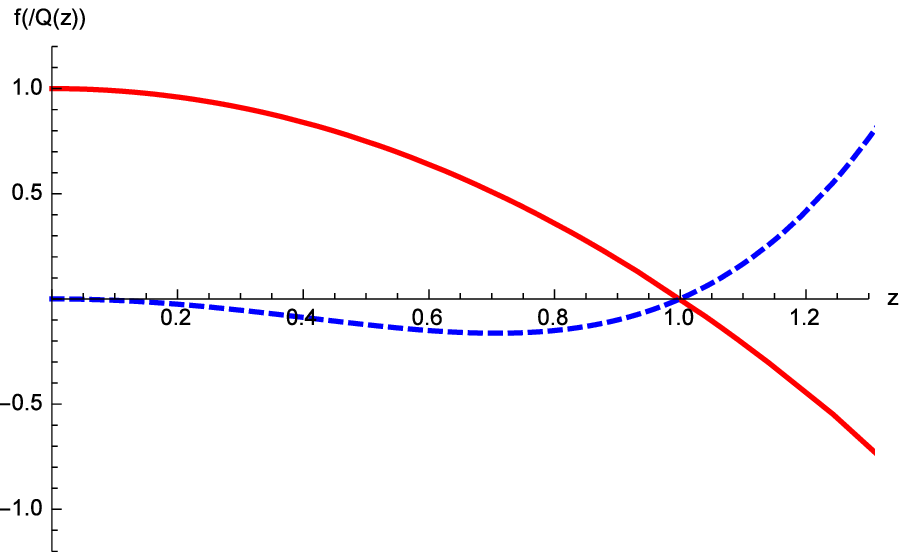}  \vskip -0.05cm \hskip 0.15 cm
\textbf{( a ) } \hskip 4.5 cm \textbf{( b )}\hskip 4.5 cm  \textbf{( c )}\\
\end{center}
\caption{Characteristic solutions when $V(\phi )= -{12\over L^2}-{16 \phi ^2\over 3L^2}+{\nu_4\phi ^4\over L^2}$ with $\nu_4=-8$. To get these solutions, we have taken $p_{22}=0,f_4=-0.001,p_2=0.0203818...$. In Panel.(a) and Panel.(b), the solutions of $\phi$ and $A_e$ are given. In Panel.(c), the solutions of $f$ is shown in red solid line, while the corresponding $Q(z)$ is shown in blue dashed line.} \label{massive-1}
\end{figure}

In this subsection, we numerically solve the gravity setup with potential like $V(\phi)=\frac{1}{L^2}\left(-12-\frac{16 }{3}\phi^2+v_4 \phi^4 \right)$. Here we set the mass of scalar to be $m^2=-\frac{16}{3L^2}$ which corresponds to dimension-2 operators in 4D. In terms of AdS dictionary, the dual operator $O_2$ is related to glueball operator. In this case, one have set $p_{22}=0$ to find solution and the $p_{22}$ corresponds to source of dual operator $O_2$ in terms of AdS/CFT. Like in the massless case, the two UV integral constants  left $p_2, f_4$ are dependent on each other due to the IR regular boundary condition $Q(z)=0$. As an example, we take $\nu_4=-8$ and $f_4=-0.001$ as an example to show the numerical process. Firstly, we take $p_2=0.015$ and $p_2=0.03$ as tests. The results are shown in Fig.\ref{massive-11}. From Fig.\ref{massive-11}.(c), we see that the behavior of $Q(z)$ near $f=0$ region are contrary to each other. Thus, the accurate value of $p_2$ should locate in between $0.015$ and $0.03$. Repeating the process, we can reduce the range of $p_2$. Finally, we find that when $p_2=0.0203818...$ all the quantities can go through the horizon smoothly, as shown in Fig.\ref{massive-1}.

\end{document}